\documentclass[aps,prb,longbibliography,superscriptaddress,twocolumn,10pt]{revtex4-2}
\usepackage{graphicx}
\usepackage{mathtools,amsmath,amssymb}
\usepackage{physics}
\usepackage{here}
\usepackage{multirow}
\usepackage{xcolor}
\usepackage{hyperref}
\usepackage{bm}
\usepackage{orcidlink}

\def\be#1\ee{\begin{align}#1\end{align}}
\newcommand{\rmd}{\mathrm{d}}

\begin{document}
\title{Light-induced Magnetization by Quantum Geometry}

\author{Hiroki Yoshida\,\orcidlink{0000-0002-4505-047X}}
\affiliation{Department of Physics, Institute of Science Tokyo, 2-12-1 Ookayama, Meguro-ku, Tokyo 152-8551, Japan}
\author{Takehito Yokoyama\,\orcidlink{0000-0003-3496-0125}}
\affiliation{Department of Physics, Institute of Science Tokyo, 2-12-1 Ookayama, Meguro-ku, Tokyo 152-8551, Japan}

\date{\today}

\begin{abstract}
    We propose a mechanism for the inverse Faraday and the inverse Cotton--Mouton effects arising from quantum geometry, characterized by the quantum metric quadrupole and the weighted quantum metric. Within a semiclassical framework based on the Boltzmann transport theory, we establish a general formalism describing light-induced magnetization in electronic systems as a second-order response to the electric field of light. Using continuum and tight-binding models, we discuss the symmetry constraints on these effects and estimate the magnitudes of the resulting magnetizations. Our results highlight a direct manifestation of quantum-geometric quantities in nonlinear magneto-optical responses and suggest a viable pathway for experimental detection.
\end{abstract}

\maketitle

\section{Introduction}
\label{sec:intro}

The Faraday effect is one of the most prominent magneto-optical phenomena, characterized by the rotation of the polarization plane of light as it propagates through a magnetized medium~\cite{Faraday_1846}. Its counterpart, the inverse Faraday effect (IFE), refers to the generation of magnetization in a material in response to circularly polarized light and has been both theoretically predicted~\cite{Pitaevskii_1961,Pershan_1963,Hertel_2006_IFE,Taguchi_Tatara_2011_IFE,Battiato_2014_QIFE,Sharma_Balatsky_2024_IFE_metal,cardosoOrbitalInverseFaraday2025,Tazuke_2025_IFE_Multiorbital} and experimentally confirmed~\cite{Zeil_Optically_1965,kimelUltrafastNonthermalControl2005,Kirilyuk_RevModPhys}. Similarly, another form of light-induced magnetization is known as the inverse Cotton--Mouton effect, in which linearly polarized light generates magnetization; this effect has also been theoretically predicted~\cite{1966_Pershan} and experimentally verified~\cite{1987_Zon}. These phenomena have attracted considerable attention as mechanisms for the optical manipulation of magnetic order, or, more broadly, for imprinting information onto quantum states using light~\cite{2025_Yeh_Quantum_Printing_I,2025_Yeh_Quantum_Printing_II,2025_Aeppli_Quantum_Printing}.

One of the most intriguing aspects of quantum states is their quantum geometry~\cite{provostRiemannianStructureManifolds1980,yuQuantumGeometryQuantum2025}. The Berry curvature~\cite{berryQuantalPhaseFactors1984,SimonHolonomy1983,AharonovAnandanPhase_1987}, which corresponds to the imaginary part of the quantum geometric tensor~\cite{Fubini1904,studyKuerzesteWegeIm1905}, is known to play a pivotal role in phenomena such as the anomalous velocity of electron wave-packets~\cite{ChagnNiu1996,Sundaram_1999_Wave-packet} and the properties of topological materials. The real part of the quantum geometric tensor is known as the quantum metric. Recent studies have shown that the quantum metric contributes to Hall viscosity~\cite{AvronViscosity1995,ReadHallviscosity2011}, the Drude weight~\cite{restaDrudeWeightSuperconducting2018}, and the superfluid weight~\cite{peottaSuperfluidityTopologicallyNontrivial2015,julkuGeometricOriginSuperfluidity2016,liangBandGeometryBerry2017,huGeometricConventionalContribution2019,julkuSuperfluidWeightBerezinskiiKosterlitzThouless2020,tormaSuperconductivitySuperfluidityQuantum2022,tianEvidenceDiracFlat2023}. Moreover, the dipole moments of the Berry curvature (Berry curvature dipole) and the quantum metric (quantum metric dipole) are known to give rise to nonlinear Hall effects~\cite{Fu_2015,Liu_2021,Kamal_2023,Gao_Science_2023}. In addition to transport phenomena, quantum geometry also manifests itself in nonlinear optical responses. A paradigmatic example is the bulk photovoltaic effect~\cite{kraut.vonbaltz1979a,belinicher.sturman1980,vonbaltz.kraut1981a,aversa.sipe1995a,sipe.shkrebtii2000,fridkin2001}, in which a dc photocurrent is generated as a second-order response to the electric field, and the corresponding nonlinear conductivity can be expressed in terms of geometric quantities~\cite{morimoto.nagaosa2016,morimoto.nagaosa2016b,ahnLowFrequencyDivergenceQuantum2020a,ma.etal2021,ahn.etal2022}. Although these studies have revealed close connections between optical responses and quantum geometry, the relationship between optically generated magnetization and quantum geometry remains unclear.

\begin{figure}
    \includegraphics[width = \columnwidth]{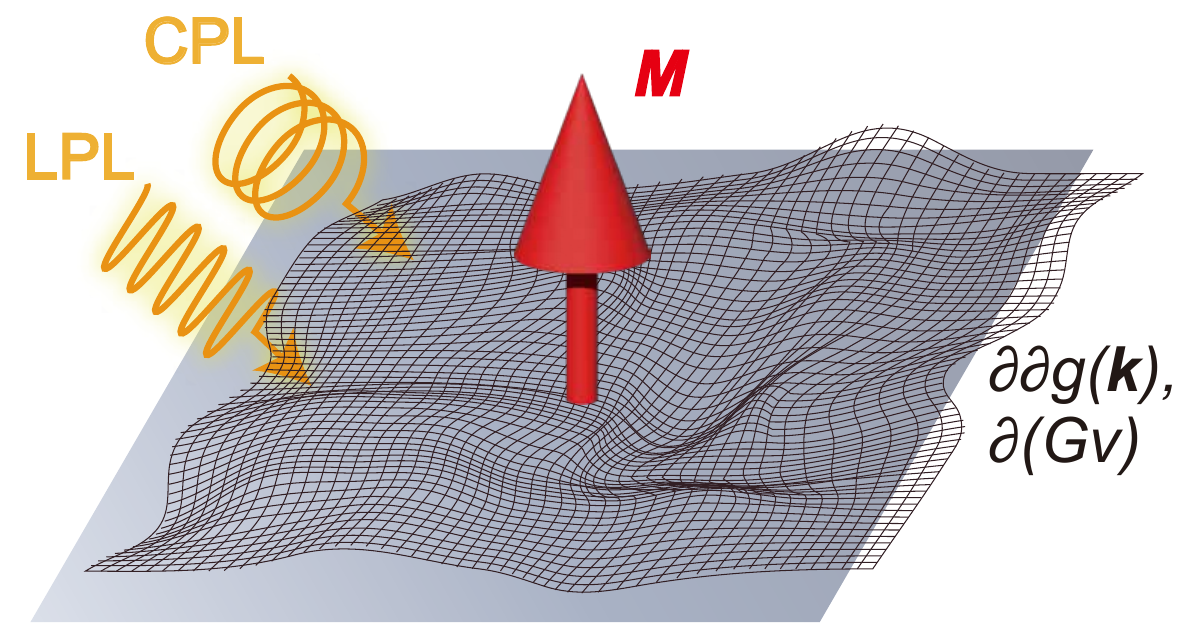}
    \caption{Schematic illustration of light-induced magnetization by quantum geometry. In response to the linearly polarized light (LPL) and circularly polarized light (CPL), the magnetization $\bm{M}$ is induced by the quantum metric quadrupole density $\partial\partial g$ and the weighted quantum metric term $\partial\left(Gv\right)$.}
    \label{fig:Schematic}
\end{figure}

In this paper, we unveil a mechanism for light-induced magnetization in electronic systems mediated by quantum-geometric effects. We consider spatially nonuniform light, characterized by an electric field with explicit position dependence. Such spatial variation modifies the semiclassical equations of motion for electrons~\cite{Lapa_Hughes_2019}, which in turn alters the second-order current response obtained from the Boltzmann transport equation. The resulting current, arising from the spatial inhomogeneity of the electric field, can be interpreted as a magnetization current, enabling us to derive an expression for light-induced magnetization as a second-order response to the electric field. We find that the induced magnetization originates from two contributions: one associated with the quantum metric quadrupole~\cite{Liu_2025_metric_quadrupole} and the other related to the weighted quantum metric~\cite{jain_anomalous_2025,YoshidaYokoyama_EG,ren_analogue_2025,ren_momentum-space_2025}. The latter contribution does not appear in transport currents but is essential for the calculation of the magnetization current. We further show that, in general, these two contributions give rise to magnetization under both linearly polarized light (LPL) and circularly polarized light (CPL), and are expected to be present in a wide range of materials (See Fig.~\ref{fig:Schematic} for schematic illustration of light-induced magnetization by quantum geometry).

This paper is organized as follows. In Sec.~\ref{sec:theory}, we provide a general theory of light-induced magnetization using the Boltzmann equation and derive general formulae for both LPL and CPL. Then, in Sec.~\ref{sec:model_calculation} we introduce continuum and lattice models and numerically evaluate the size of the induced magnetization. We conclude this paper in Sec.~\ref{sec:concl}.

\section{General theory}
\label{sec:theory}

In this section, we develop a general theoretical framework for light-induced magnetization. We calculate the magnetization current induced by light using the semiclassical equations of motion for electrons combined with the Boltzmann transport theory. We show that the magnetization, as a second-order response to the electric field, can be expressed in terms of quantum-geometric quantities.

\subsection{Equation of motion}
\label{subsec:EOM}
We begin with the semiclassical equations of motion for an electron in a spatially nonuniform electric field. Throughout this paper, we neglect the magnetic-field component of the light and focus exclusively on the electric field. The oscillating electric field is expressed in terms of a complex field $\tilde{E}$ as
\be
    E_{a}(\bm{r},t)&=\Re\left[\tilde{E}_{a}(\bm{r})e^{i\omega t}\right]\quad\left(\tilde{E}_{a}\in\mathbb{C}\right),
\ee
where $a=x,y,z$ and the electric field is assumed to be oscillating in a single frequency $\omega$. Using the scalar potential $\phi(\bm{r},t)$ and vector potential $\bm{A}(\bm{r},t)$, this electric field can be written as
\be
    \bm{E}(\bm{r},t)&=-\nabla\phi(\bm{r},t)-\frac{\partial\bm{A}(\bm{r},t)}{\partial t}.
\ee
The Hamiltonian of the system under this electric field is
\be
    \hat{H}&=\hat{H}_0\left[\hat{\bm{q}}-Q\bm{A}(\hat{\bm{r}},t)\right]+Q\phi(\hat{\bm{r}},t),
\ee
where $\hat{H}_0(\hat{\bm{q}})$ is a static Hamiltonian including a periodic potential and $Q=-e$ is the charge of an electron. The semiclassical wave-packet dynamics of a system with this Hamiltonian is studied in Refs.~\cite{Sundaram_1999_Wave-packet,GaoXiao_2019_NonreciprocalDirectional}. The semiclassical equations of motion is given as
\be
    \left\{\begin{array}{l}
        \dot{r}_a=\frac{1}{\hbar}\frac{\partial\varepsilon'_n}{\partial q_a}-\Omega'^{qq}_{ab}\dot{q}_b-\Omega'^{qr}_{ab}\dot{r}_b\\
        \dot{q}_a=\frac{Q}{\hbar}E_a(\bm{r},t)+\Omega'^{rq}_{ab}\dot{q}_b,     
    \end{array}\right.\label{eq:EOM}
\ee
where $\varepsilon'_n$ is the energy of an $n$th band including the effect of an external electric field, $\Omega'^{qq}$ and $\Omega'^{qr}$ are the Berry curvatures of the $n$th band defined as
\be
    \Omega'^{qq}_{ab}&\coloneqq -2\Im\left[\left\langle\partial_{q_a}u_n'\middle|\partial_{q_b}u_n'\right\rangle\right],\label{eq:def_Omegaqq}\\
    \Omega'^{qr}_{ab}&\coloneqq -2\Im\left[\left\langle\partial_{q_a}u_n'\middle|\partial_{r_b}u_n'\right\rangle\right].\label{eq:def_Omegaqr}
\ee
Here, $\Omega'^{qr}_{ab}=-\Omega'^{rq}_{ba}$ and these quantities are defined using the perturbed state
\be
    \left\lvert u_n'(\bm{r},\bm{q})\right\rangle &=|u_n\rangle+\sum_{m\neq n}\frac{QE_a(\bm{r},t)(A_{mn})_a}{\varepsilon_m-\varepsilon_n}|u_m\rangle,\label{eq:PerturbedState}
\ee
where $|u_n\rangle$ is the unperturbed periodic part of Bloch state and $(A_{mn})_a=i\langle u_m|\partial_{q_a}u_n\rangle$ is the Berry connection.

Below, we evaluate each quantity appearing in Eq.~\eqref{eq:EOM}. As will be shown in the following sections, to achieve our final goal of deriving the magnetization response proportional to $\tilde{E}\tilde{E}^*$, it is sufficient to retain terms up to the order of $\tilde{E}\partial_r\tilde{E}^*$ in the equations of motion. We begin by computing the mixed Berry curvature $\Omega'^{qr}$. Since the spatial derivative in Eq.~\eqref{eq:def_Omegaqr} acts on the electric field in Eq.~\eqref{eq:PerturbedState}, we obtain, without further approximations,
\begin{widetext}
    \be
        \Omega'^{qr}_{ab}&=-\frac{Q}{2\hbar}G_{aj}\partial_{r_b}E_j-2Q^2\Im\left[\sum_{m\neq n}\partial_{q_a}\left(\frac{\left(A_{nm}\right)_i}{\varepsilon_m-\varepsilon_n}\right)\frac{\left(A_{mn}\right)_j}{\varepsilon_m-\varepsilon_n}+i\sum_{\substack{m\neq n\\l\neq n}}\frac{\left(A_{nm}\right)_i\left(A_{ml}\right)_a\left(A_{ln}\right)_j}{(\varepsilon_m-\varepsilon_n)(\varepsilon_l-\varepsilon_n)}\right]E_i\partial_{r_b}E_j,\label{eq:omegaqr_explicit}
    \ee
\end{widetext}
where
\be
    G_{aj}&\coloneqq 4\hbar\Re\left[\sum_{m\neq n}\frac{\left(A_{nm}\right)_a\left(A_{mn}\right)_j}{\varepsilon_m-\varepsilon_n}\right]
\ee
is a quantity called weighted quantum metric~\cite{jain_anomalous_2025,YoshidaYokoyama_EG}. This quantity appears in electric polarization induced by a uniform electric field~\cite{Gao_positional_2014} and nonadiabatic dynamics of electrons~\cite{ren_momentum-space_2025,YoshidaYokoyama_EG,ren_analogue_2025}. Then, by recursive substitution (see Appendix~\ref{app:EOM} for detail), we obtain
\be
    \dot{q}_a&\approx\frac{Q}{\hbar}E_a+\frac{1}{2}\left(\frac{Q}{\hbar}\right)^2G_{bj}E_b\partial_{r_a}E_j.\label{eq:q_EOM_EdE}
\ee
We now turn to the calculation of $\dot{r}_a$. The correction to the group velocity can be obtained by evaluating the expectation value of the energy of the wave-packet~\cite{GaoXiao_2019_NonreciprocalDirectional}. The full expression for $\dot{r}$, including terms up to the order of $\tilde{E}\partial_{r}\tilde{E}^*$ and contributions from $\Omega'^{qq}$, is presented in Appendix~\ref{app:EOM}. On the other hand, for the purpose of calculating the current up to the order of $\tilde{E}\partial_{r}\tilde{E}^*$, it is sufficient to retain $\dot{r}$ only up to the order of $\partial_r\tilde{E}$, since the distribution function in metals contains at least first order in $\tilde{E}$, as discussed in Sec.~\ref{subsec:Current}. In this case, the calculation follows the same procedure as in Ref.~\cite{GaoXiao_2019_NonreciprocalDirectional}. The resulting expression is
\be
    \dot{r}_a&\approx v_a-\frac{Q}{\hbar}\Omega^{qq}_{ab}E_b+\frac{Q}{2\hbar}\left(-\partial_{q_a}g_{bj}+G_{aj}v_b\right)\partial_{r_b}E_j,\label{eq:r_EOM_dE}
\ee
where $v_a=\partial_{q_a}\varepsilon_n/\hbar$ is the group velocity of an electron in the $n$th band, the Berry curvature $\Omega^{qq}$ is defined in terms of the unperturbed states, and $g_{ij}\coloneqq \Re\left[\sum_{m\neq n}\left(A_{nm}\right)_i\left(A_{mn}\right)_j\right]$ is the quantity known as the quantum metric. The derivative of this quantity is referred to as the quantum metric dipole. The term containing the quantum metric dipole in Eq.~\eqref{eq:r_EOM_dE} originates from the correction to the group velocity, $\partial_{q_a}\varepsilon_n'/\hbar$, in Eq.~\eqref{eq:EOM}, as also discussed in Refs.~\cite{Lapa_Hughes_2019,GaoXiao_2019_NonreciprocalDirectional}.

\subsection{Electric current}
\label{subsec:Current}
We next calculate the electric current induced by $E\partial_r E$. Since we focus on phenomena that do not vanish upon time averaging, we consider the direct current proportional to $\tilde{E}\partial_r\tilde{E}^*$.

The electric current flowing in the $a$ direction can be calculated as 
\be
    j_a&=Q\int \left[\rmd \bm{q}\right] D\dot{r}_af(\varepsilon_n(\bm{q})),\label{eq:def_current}
\ee
where $\left[\mathrm{d}\bm{q}\right] = \frac{\mathrm{d}\bm{k}}{(2\pi)^d}$ with $d$ denoting the dimensionality of the system, and $D$ is the modified density of states~\cite{XiaoShiNiu_2005_DoS}, given by
\be
    D=1+\tr\Omega'^{qr}.\label{eq:DOS}
\ee
Here, $f(\varepsilon_n(\bm{q}))$ is the distribution function. The nonequilibrium distribution function can be obtained from the Boltzmann transport equation within the relaxation-time approximation,
\be
    \pdv{f}{t}+\dot{\bm{q}}\cdot\pdv{f}{\bm{q}}=-\frac{f-f_0}{\tau}\label{eq:Boltzmann},
\ee
where $f_0$ is the equilibrium distribution function and $\tau$ denotes the relaxation time.

We derive the distribution function from Eq.~\eqref{eq:Boltzmann} using $\dot{q}$ that is also obtained (Eq.~\eqref{eq:q_EOM_EdE}) in the previous section. We write $f$ as a Fourier series of complex functions $\tilde{f}_m$: $f(\varepsilon_n(\bm{q}))=\Re\left[\sum_{m=0}^{\infty}\tilde{f}_m(\varepsilon_n(\bm{q}))e^{im\omega t}\right]$. By substituting this into Eq.~\eqref{eq:Boltzmann}, we obtain each $\tilde{f}_m$. We note that, $\dot{q}_a$ includes a term proportional to $E\partial E$ and thus, there is a term proportional to $E\partial E$ in the distribution function, different from calculations in previous studies (Ref.~\cite{Fu_2015}, for example). We show the details of the calculation in Appendix~\ref{app:distribution function}. The results up to $E\partial_rE$ are given by 
\be
    \tilde{f}_0&\approx f_0+\frac{\tau}{4}\left(\frac{Q}{\hbar}\right)^2G_{bj}\frac{\partial f_0}{\partial q_a}\tilde{E}_b\partial_{r_a}\tilde{E}_j^*,\label{eq:f0}\\
    \tilde{f}_1&\approx \frac{-1}{1+i\tau\omega}\left(\frac{\tau Q}{\hbar}\right)\frac{\partial f_0}{\partial q_a}\tilde{E}_a,\label{eq:f1}\\
    \tilde{f}_2&\approx \frac{1}{1+2i\tau\omega}\frac{\tau}{4}\left(\frac{Q}{\hbar}\right)^2G_{bj}\frac{\partial f_0}{\partial q_a}\tilde{E}_b\partial_{r_a}\tilde{E}_j,\\
    \tilde{f}_m&\approx 0\qquad (m\geq3).\label{eq:fn}
\ee
In this paper, we focus on metallic systems. Therefore, for our calculation of the current Eq.~\eqref{eq:def_current} we only use Fermi surface terms. Also, as we are interested in dc response, what we use in the following are $\tilde{f}_1\,\left(\propto \tilde{E}\right)$ and the second term in $\tilde{f}_0\,\left(\propto \tilde{E}\partial_r\tilde{E}^*\right)$.

By substituting Eqs.~\eqref{eq:omegaqr_explicit}, \eqref{eq:q_EOM_EdE}, \eqref{eq:r_EOM_dE}, \eqref{eq:DOS}, \eqref{eq:f0}, and \eqref{eq:f1} to Eq.~\eqref{eq:def_current}, we obtain a direct current proportional to $\tilde{E}\partial\tilde{E}^*$ as
\begin{widetext}
    \be
        j^0_a&=\frac{\tau Q^3}{8\hbar^2}\int\left[\rmd\bm{q}\right]\left[G_{bj}v_a\frac{\partial f_0}{\partial q_i}+\frac{1}{1+i\tau\omega}\left(\partial_{q_a}g_{ij}+G_{ij}v_a-G_{aj}v_i\right)\frac{\partial f_0}{\partial q_b}\right]\tilde{E}_b\partial_{r_i}\tilde{E}_j^*+(\mathrm{c.c.}),\label{eq:result_current_EdE}
    \ee
\end{widetext}
where $c.c.$ represents the complex conjugate of the first term.

\subsection{Magnetization}
\label{subsec:magnetization}

We derive the magnetization induced by light using the obtained expression for the electric current in Eq.~\eqref{eq:result_current_EdE}.

The magnetization current $\bm{j}^M$ is defined as the curl of a position-dependent magnetization $\bm{M}$, 
\be
    \bm{j}^M&=\nabla_r\times\bm{M},
\ee
where $\nabla_r$  denotes the derivative with respect to position. We focus on a static magnetization that remains nonzero after time averaging. Since the electric field is oscillatory, such a static magnetization should be at least second order in the electric field and can be written as
\be
    \tilde{M}_a^0&=\sigma_{abc}\tilde{E}_b\tilde{E}^*_c,\label{eq:def_sigma}
\ee
where $M_a(\bm{r},t)=\Re\left[\sum_{m=0}^{\infty}\tilde{M}^m_ae^{im\omega t}\right]$. By taking the curl of the nonoscillating component of the magnetization, $M^0_a:=\Re\left[\tilde{M}^0_a\right]$, we obtain the static magnetization current $\tilde{j}^{M,0}_a=\varepsilon_{aic}\left(\sigma_{cbj}+\sigma^*_{cjb}\right)\tilde{E}_b\partial_{r_i}\tilde{E}^*_j+\mathrm{c.c.}$. By comparing this expression with Eq.~\eqref{eq:result_current_EdE}, we identify $\sigma$ in Eq.~\eqref{eq:def_sigma} as
\begin{widetext}
    \be
        \sigma_{abc}+\sigma^*_{acb}&=\frac{\tau Q^3}{16\hbar^2\left(1+i\tau\omega\right)}\int\left[\rmd\bm{q}\right]\varepsilon_{aij}\left(\partial_{q_i}g_{jc}+G_{jc}v_i-G_{ic}v_j\right)\frac{\partial f_0}{\partial q_b}.\label{eq:result_sigma}
    \ee
\end{widetext}
This result is, coincidentally, analogous to the expression of the current proportional to $\partial_r E$ in insulator obtained in Ref.~\cite{GaoXiao_2019_NonreciprocalDirectional}. The first term originates from the correction to the energy, whereas the latter two terms in the brackets result from the correction to the state (Eq.~\eqref{eq:PerturbedState}). For the transport current, only the first term contributes~\cite{Lapa_Hughes_2019,GaoXiao_2019_NonreciprocalDirectional}, and the magnetization current is discarded. In contrast, in our case, the latter two terms play an essential role.

We now return to our original motivation of deriving the magnetization induced by light and consider its dependence on the polarization of the light. Physically, the magnetization is obtained by taking the real part of Eq.~\eqref{eq:def_sigma}. For LPL, $\operatorname{Re}\left[\tilde{E}_b\tilde{E}^*_c\right]$ is nonzero and symmetric under the exchange of indices $b$ and $c$. In contrast, for CPL, $\operatorname{Im}\left[\tilde{E}_b\tilde{E}^*_c\right]$ is nonzero and antisymmetric with respect to the indices (see Appendix~\ref{app:polarization}). Accordingly, we define $\sigma^{\mathrm{LPL}}$ for LPL as
\be
    \sigma^{\mathrm{LPL}}_{abc}&\coloneqq \operatorname{Re}\left[\frac{\sigma_{abc}+\sigma_{acb}}{2}\right]=\sigma^{\mathrm{LPL}}_{acb},
\ee
and $\sigma^{\mathrm{CPL}}$ for CPL as
\be
    \sigma^{\mathrm{CPL}}_{abc}&\coloneqq -\operatorname{Im}\left[\frac{\sigma_{abc}-\sigma_{acb}}{2}\right]=-\sigma^{\mathrm{CPL}}_{acb}.
\ee
For a general polarization of light, both coefficients contribute to the observable magnetization. By performing integration by parts, these coefficients can be written as
\begin{widetext}
    \be
        \sigma^{\mathrm{LPL}}_{abc}&=\frac{-\tau Q^3}{64\hbar^2\left(1+\tau^2\omega^2\right)}\int\left[\rmd\bm{q}\right]\varepsilon_{aij}\left[\partial_{q_b}\partial_{q_i}g_{jc}+2\partial_{q_b}\left(G_{jc}v_i\right)+\left(b\leftrightarrow c\right)\right]f_0,\label{eq:sigma_LPL}\\
        \sigma^{\mathrm{CPL}}_{abc}&=\frac{-\tau^2\omega Q^3}{64\hbar^2\left(1+\tau^2\omega^2\right)}\int\left[\rmd\bm{q}\right]\varepsilon_{aij}\left[\partial_{q_b}\partial_{q_i}g_{jc}+2\partial_{q_b}\left(G_{jc}v_i\right)-\left(b\leftrightarrow c\right)\right]f_0.\label{eq:sigma_CPL}
    \ee
\end{widetext}
This is our main result.

The coefficient $\sigma^{\mathrm{LPL}}$ represents the magnetization induced by LPL, corresponding to the inverse Cotton--Mouton effect, while $\sigma^{\mathrm{CPL}}$ describes the magnetization induced by CPL, corresponding to the inverse Faraday effect. In both cases, the effects are governed by a quadrupole-like quantity of the quantum metric $\int[\mathrm{d}\bm{k}]f_0\partial \partial g$, which we refer to as the \textit{quantum metric quadrupole}. This quantity has previously been studied in the context of third-order electric responses~\cite{Liu_2022_BCP,Liu_2025_metric_quadrupole}. In addition to the quantum metric quadrupole, these responses also include terms involving the weighted quantum metric $G$. These terms originate from perturbations to the Bloch states and do not appear in the transport current~\cite{GaoXiao_2019_NonreciprocalDirectional,Lapa_Hughes_2019}.

\section{Model Calculation}
\label{sec:model_calculation}

In this section, we calculate $\sigma^{\mathrm{LPL}}$ and $\sigma^{\mathrm{CPL}}$ for two-band models and evaluate their magnitudes with an eye toward experimental detection. In the following, we focus on two-dimensional materials. We first introduce an effective continuum model to elucidate the general features of the induced magnetization. We then consider a lattice model and estimate the size of the effects.

Before introducing the specific models, we discuss the role of symmetries in two-dimensional systems. Here, we summarize the main results, while the detailed analysis is provided in Appendix~\ref{app:sym}. For a response of the form given in Eq.~\eqref{eq:def_sigma}, the simultaneous presence of symmetries under a mirror operation with respect to a mirror plane perpendicular to the system $m_{\perp}$ and a $n$-fold rotational symmetry around an axis perpendicular to the system $C_{nz}$ with $n\neq 2$ enforces $\sigma_{zab}^{\mathrm{LPL}}=0$. In contrast, neither mirror nor rotational symmetries constrain the response to CPL. Therefore, in order to observe responses to both LPL and CPL, the system should break either mirror or rotational symmetries.

\subsection{Continuum model}
We consider a two-dimensional massive Dirac model with anisotropic quadratic corrections, described by
\be
    H(q_x,q_y)&=\hbar v\left[\left(q_x+\alpha q_x^2\right)\sigma_x+\left(q_y+\beta q_y^2\right)\sigma_y\right]+M\sigma_z,\label{eq:2nd_Dirac_model_Hamiltonian}
\ee
where $\sigma_i\ (i=x,y,z)$ are Pauli matrices acting on a two-component internal space, and $M$ is the mass-gap parameter.
We have presented expressions of quantum metric quadrupole and weighted quantum
 metric terms in Eqs.~\eqref{eq:sigma_LPL} and \eqref{eq:sigma_CPL} for general two-level systems in Appendix~\ref{app:QMQ}, which we apply to our model Hamiltonian here.
For $\alpha=\beta$, we find $\sigma_{zaa}=0$ due to the mirror symmetry (see Appendix~\ref{app:sym}) and $\sigma^{\mathrm{LPL}}_{zxy}=0$ by the direct calculation. Hence, we here focus on the case with $\alpha \ne \beta$.

It should be also noted that the electric current transforms as a polar vector under spatial symmetry operations. For mirror reflections with respect to planes normal to the $x$ and $y$ directions, denoted by $m_{100}$ and $m_{010}$, respectively, the corresponding representations are given by
\be
    U_{m_{100}} = i\sigma_y,\qquad U_{m_{010}} = i\sigma_x.
\ee
In the absence of quadratic corrections $(\alpha=\beta=0)$, the Hamiltonian is isotropic in momentum space and thus possesses a continuous rotational symmetry $C_{\infty z}$. Since this isotropic Dirac model has both mirror $m_{\perp}$ and rotation $C_{\infty z}$ symmetries, the response to LPL is forbidden.

One may also construct a model with an anisotropic Dirac cone without quadratic terms, $H=\alpha_x q_x\sigma_x+\alpha_y q_y\sigma_y+M\sigma_z$ with $\alpha_x\neq\alpha_y$. Since this system breaks rotation $C_{\infty z}$ symmetries, one would then expect a nonzero response $\sigma^{\mathrm{LPL}}_{zxy}$. However, as shown by explicit calculation in Appendix~\ref{app:anisotropic_massive_dirac}, this model is still too simple: the contributions cancel upon momentum integration, resulting in $\sigma^{\mathrm{LPL}}_{zxy}=0$. Similarly, for a system described by the Hamiltonian $H=\alpha_x q_x^2\sigma_x+\alpha_y q_y^2\sigma_y+M\sigma_z$, the response $\sigma^{\mathrm{LPL}}$ also vanishes. Therefore, Eq.~\eqref{eq:2nd_Dirac_model_Hamiltonian} constitutes the minimal model that yields nonvanishing responses to both LPL and CPL.

\begin{figure}
    \includegraphics[width=\columnwidth]{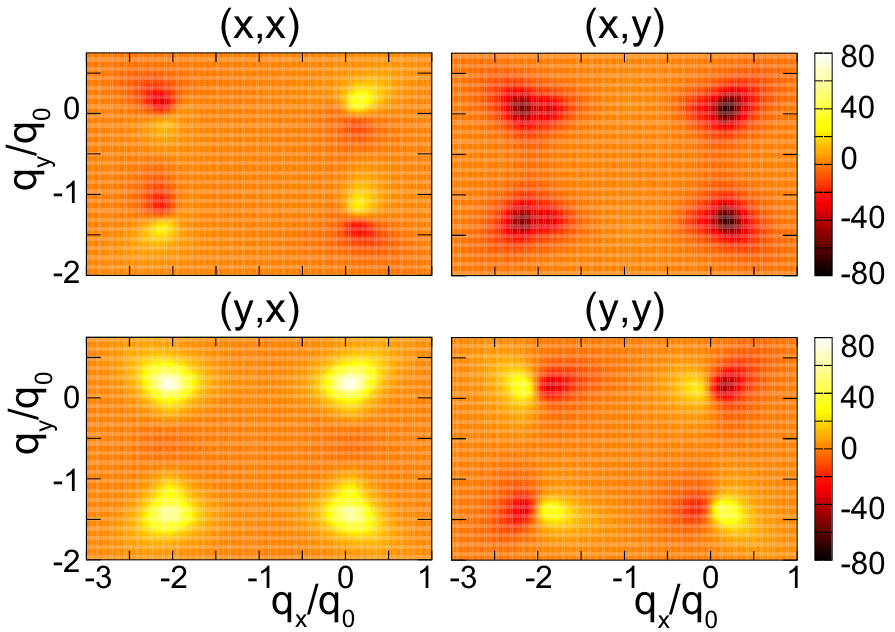}
    \caption{The quantum metric quadrupole density term $(q_0^4\sum_{i,j}\varepsilon_{zij}\partial_a\partial_ig_{jb})$ in the momentum space. The labels above each panel indicate the corresponding $(a,b)$ pairs. Parameters are $\alpha q_0=0.5,\,\beta q_0=0.8$ and $M/(\hbar v q_0)=0.3$, where $q_0$ determines the length scale of this system.}
    \label{fig:2ndDirac_ddg}
\end{figure}
\begin{figure}
    \includegraphics[width=\columnwidth]{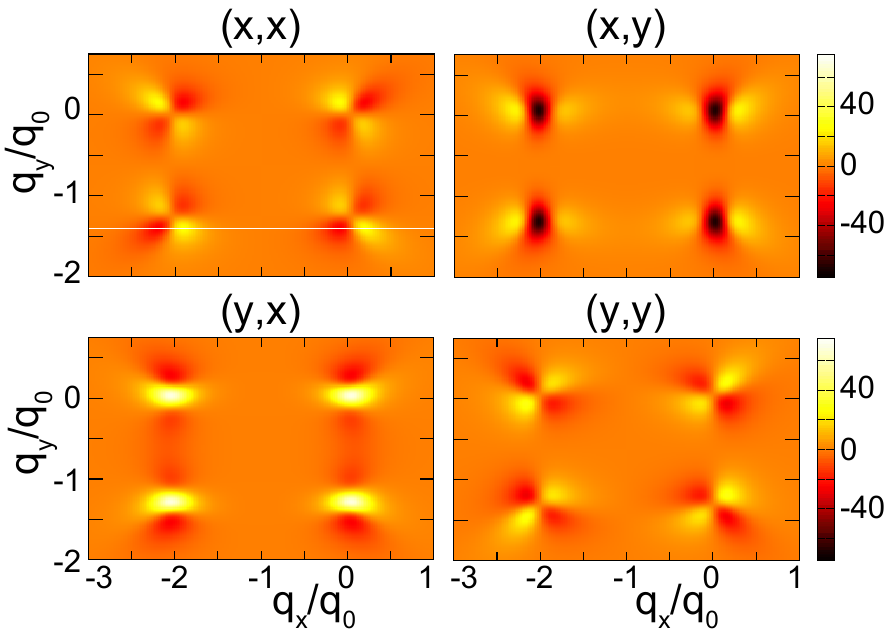}
    \caption{The weighted quantum metric density term $(q_0^4\sum_{i,j}\varepsilon_{zij}\partial_a\left(G_{jb}v_i\right))$ in the momentum space. The labels above each panel indicate the corresponding $(a,b)$ pairs. Parameters are $\alpha q_0=0.5,\,\beta q_0=0.8$ and $M/(\hbar v q_0)=0.3$, where $q_0$ determines the length scale of this system.}
    \label{fig:2ndDirac_dGv}
\end{figure}

\begin{figure}[t]
    \includegraphics[width=\columnwidth]{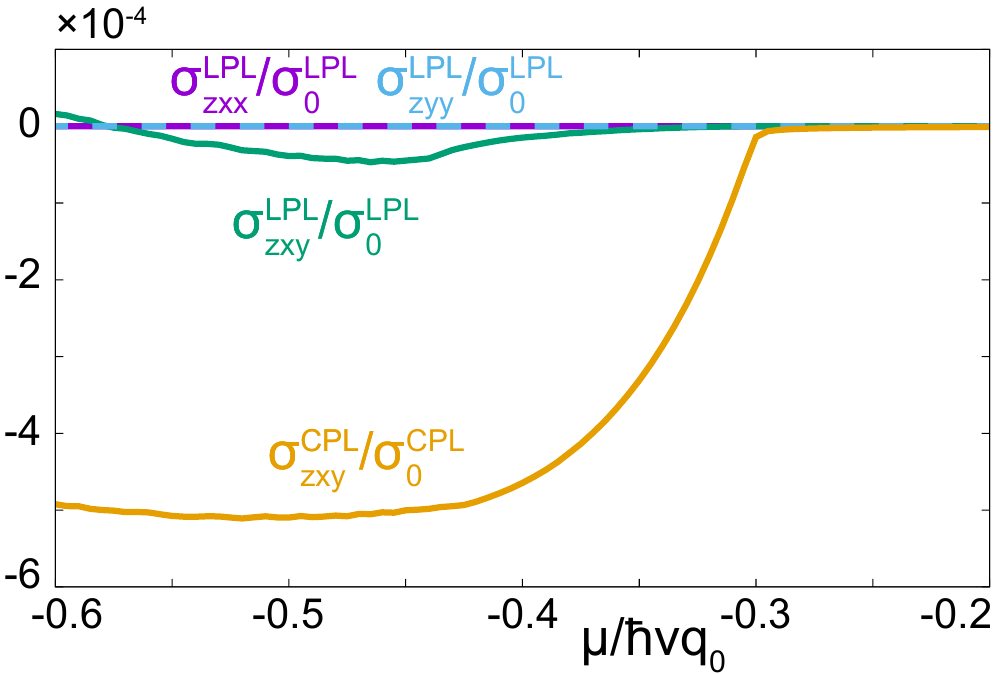}
    \caption{Calculated $\sigma^{\mathrm{LPL}}_{zab}$ and $\sigma^{\mathrm{CPL}}_{zxy}$ as a function of the chemical potential $\mu$ divided by the energy scale $\hbar vq_0$. The results are normalized by $\sigma^{\mathrm{LPL}}_0=-\tau Q^3/64q_0^2\hbar^2(1+\tau^2\omega^2)$ and $\sigma^{\mathrm{CPL}}_0=-\tau^2\omega Q^3/64q_0^2\hbar^2(1+\tau^2\omega^2)$. The parameters are chosen as $\alpha q_0=0.5,\,\beta q_0=0.8$ and $M/(\hbar v q_0)=0.3$, where $q_0$ determines the length scale of this system.}
    \label{fig:2ndDirac_sigma}
\end{figure}

We plot the quantum metric quadrupole and the weighted quantum metric terms for this model in Figs.~\ref{fig:2ndDirac_ddg} and \ref{fig:2ndDirac_dGv}, respectively. These quantities take nonzero values around four points $(q_x,q_y)=(0,0),(-1/\alpha,0),(0,-1/\beta)$, and $(-1/\alpha,-1/\beta)$, where the energy gap becomes minimal. In fact, the energies are given as
\be
    \varepsilon_{\pm}&=\pm\hbar v\sqrt{\left\{q_x(1+\alpha q_x)\right\}^2+\left\{q_y(1+\beta q_y)\right\}^2+\left(\frac{M}{\hbar v}\right)^2}.
\ee
The quantum metric quadrupole and weighted quantum metric contributions are of the same order of magnitude, highlighting the importance of including the weighted quantum metric terms, which are absent in conventional transport-current calculations. By integrating these quantities, we obtain the coefficients $\sigma$ for LPL and CPL. The results are summarized in Fig.~\ref{fig:2ndDirac_sigma}. For this model, we find nonzero responses to both LPL and CPL, corresponding to the inverse Cotton--Mouton and the inverse Faraday effects, respectively. The coefficients $\sigma^{\mathrm{LPL}}_{zxx}$ and $\sigma^{\mathrm{LPL}}_{zyy}$ vanish due to the simplicity of the model.

\subsection{Lattice model}

\begin{figure}
    \includegraphics[width=0.6\columnwidth]{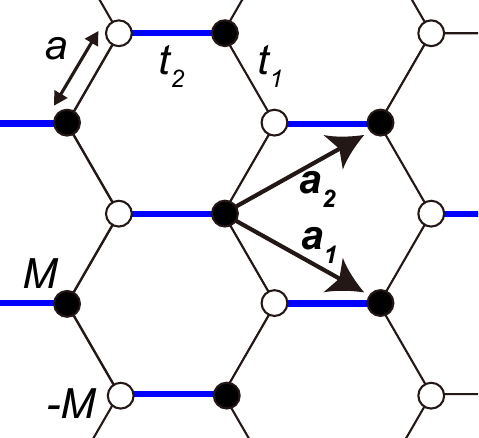}
    \caption{Schematic illustration of the tight-binding model on hexagonal lattice. Two kinds of nearest neighbor hoppings $t_1$ and $t_2$ and staggered on-site potential $M$ are introduced. Lattice constant is $a$ and primitive lattice vectors are $\bm{a}_1=\left(3a/2,-\sqrt{3}a/2\right)$ and $\bm{a}_2=\left(3a/2,\sqrt{3}a/2\right)$. Reciprocal lattice vectors are $\bm{b}_1=\left(2\pi/(3a),-2\pi/(\sqrt{3}a)\right)$ and $\bm{b}_2=\left(2\pi/(3a),2\pi/(\sqrt{3}a)\right)$.}
    \label{fig:TBM}
\end{figure}

To overcome the limitations of the continuum model and provide a realistic estimate of the magnitude of the effect, we introduce a tight-binding model on a hexagonal lattice. We consider a hexagonal lattice with nearest-neighbor hopping amplitude $t$ and a staggered on-site potential $M$, as shown in Fig.~\ref{fig:TBM}. We introduce an anisotropic hopping $t_2$ along the direction parallel to the $x$ axis, which differs from the other hopping amplitudes $t_1$. In the absence of this anisotropy, the system possesses $m_{010}$ and $C_{3z}$ symmetries, which enforce a vanishing response to LPL, as discussed in Appendix~\ref{app:sym}. When the anisotropy $t_2\neq t_1$ is introduced, the model retains only the $m_{010}$ symmetry, which enforces $\sigma_{zxx}=\sigma_{zyy}=0$ but does not constrain the other tensor components.

\begin{figure}
    \includegraphics[width=\columnwidth]{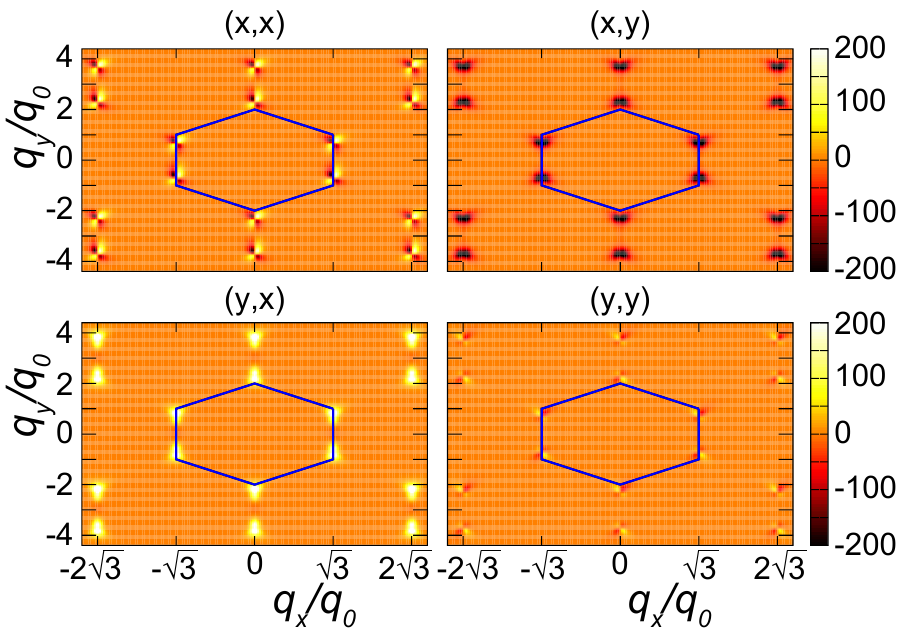}
    \caption{The quantum metric quadrupole density term $\left(q_0^4\sum_{i,j}\varepsilon_{zij}\partial_a\partial_ig_{jb}\right)$ in the momentum space for the hexagonal lattice model. The labels above each panel indicate the corresponding $(a,b)$ pairs. The blue hexagon indicates the Brillouin zone. Here, $q_0=2\pi/(3a)$. Values of the parameters are $t_2/t_1=1.5$ and $M/t_1=0.3$.}
    \label{fig:Graphene_ddg}
\end{figure}

\begin{figure}
    \includegraphics[width=\columnwidth]{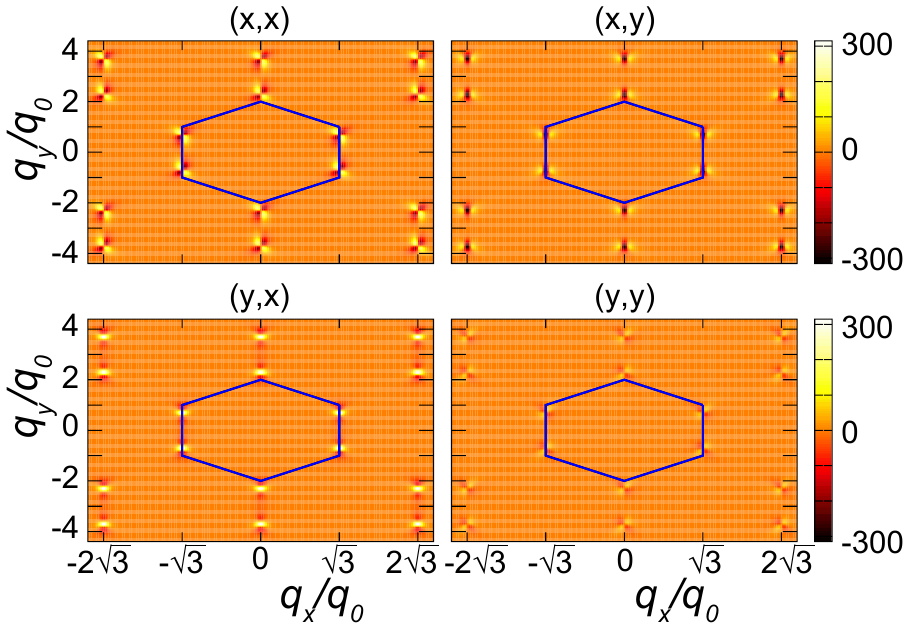}
    \caption{The weighted quantum metric term $\left(q_0^4\sum_{i,j}\varepsilon_{zij}\partial_a\left(G_{jb}v_i\right)\right)$ in the momentum space for the hexagonal lattice model. The labels above each panel indicate the corresponding $(a,b)$ pairs. The blue hexagon indicates the Brillouin zone. Here, $q_0=2\pi/(3a)$. Values of the parameters are $t_2/t_1=1.5$ and $M/t_1=0.3$.}
    \label{fig:Graphene_dGv}
\end{figure}

The tight-binding Hamiltonian of this system is given by
\be
    H(\bm{q})&=\bm{h}(\bm{q})\cdot\boldsymbol{\sigma},
\ee
where
\be
    h_x(\bm{q})&=2t_1\cos\left(\frac{\sqrt{3}}{2}aq_y\right)\cos\left(\frac{a}{2}q_x\right)+t_2\cos\left(aq_x\right),\\
    h_y(\bm{q})&=-2t_1\cos\left(\frac{\sqrt{3}}{2}aq_y\right)\sin\left(\frac{a}{2}q_x\right)+t_2\sin\left(aq_x\right),\\
    h_z(\bm{q})&=M.
\ee
Here, we use the sublattice basis, and $a$ denotes the lattice constant. The energies are obtained as 
\be
    \varepsilon_{\pm}&=\pm t_1\left[4\left\{\cos\left(\frac{\sqrt{3}a}{2}q_y\right)+\frac{t_2}{2t_1}\cos\left(\frac{3a}{2}q_x\right)\right\}^2\right.\nonumber\\
    &\hspace{1.5cm}\left.+\left\{\frac{t_2}{t_1}\sin\left(\frac{3a}{2}q_x\right)\right\}^2+\left(\frac{M}{t_1}\right)^2\right]^{\frac{1}{2}}.
\ee
Thus, this model exhibits minima of the energy gap at
$$\bm{q}=\left(0,\frac{2}{\sqrt{3}a}\cos^{-1}\left(-\frac{t_2}{2t_1}\right)\right)$$
and
$$\bm{q}=\left(\frac{2\pi}{3a},\frac{2}{\sqrt{3}a}\cos^{-1}\left(\frac{t_2}{2t_1}\right)\right).$$
When $t_1\neq t_2$, these gap minima are shifted away from the $K$ and $K'$ points. The quantum metric quadrupole and the weighted quantum metric contributions are plotted over the Brillouin zone in Figs.~\ref{fig:Graphene_ddg} and \ref{fig:Graphene_dGv}, respectively. Both quantities become nonzero in the vicinity of the gap minima and have comparable magnitudes.

\begin{figure}
    \includegraphics[width=\columnwidth]{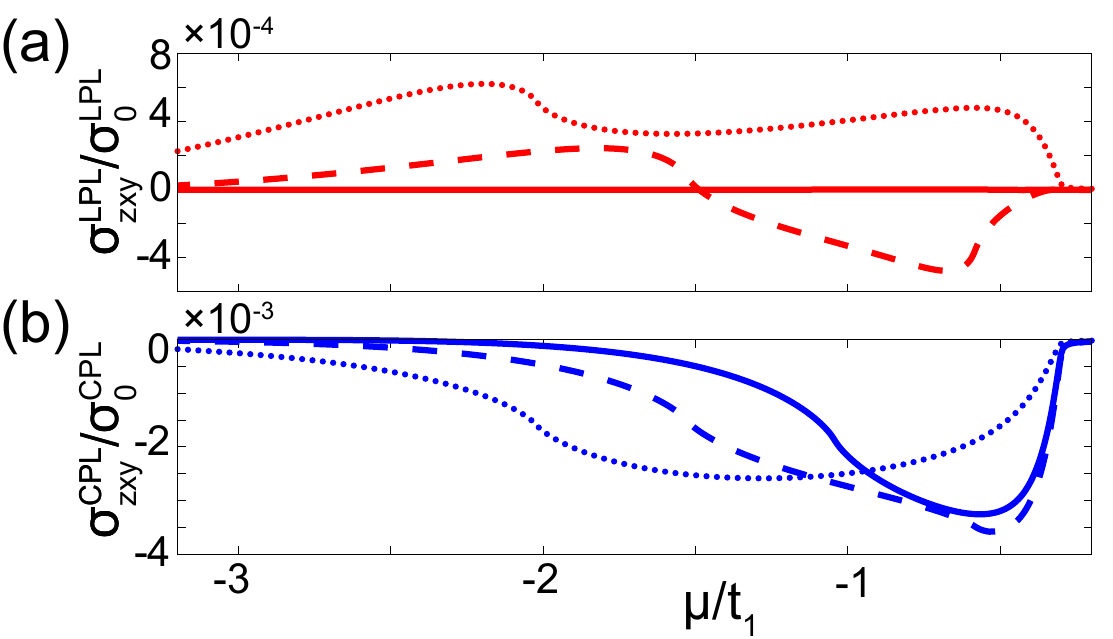}
    \caption{Calculated (a) $\sigma^{\mathrm{LPL}}_{zxy}$ and (b) $\sigma^{\mathrm{CPL}}_{zxy}$ as a function of chemical potential $\mu$ divided by $t_1$. Definitions of $\sigma^{\mathrm{LPL}}_0$ and $\sigma^{\mathrm{CPL}}_0$ are the same as those in Fig.~\ref{fig:2ndDirac_sigma} with $q_0=2\pi/(3a)$. The solid, dashed, and dotted lines represent cases with $t_2/t_1=1.0,\,1.5$, and $2.0$, respectively. We fix $M/t_1=0.3$.}
    \label{fig:Graphene_sigma}
\end{figure}

In Fig.~\ref{fig:Graphene_sigma}, we plot $\sigma^{\mathrm{LPL}}_{zxy}$ in (a) and $\sigma^{\mathrm{CPL}}_{zxy}$ in (b) as a function of the chemical potential. The solid, dashed, and dotted lines correspond to $t_2/t_1=1.0,\,1.5$, and $2.0$, respectively. By symmetry, $\sigma^{\mathrm{LPL}}_{zaa}=0$. As expected from the symmetry analysis, $\sigma^{\mathrm{LPL}}_{zxy}$ vanishes when $t_2/t_1=1.0$. For $t_2/t_1\neq 1.0$, nonzero responses emerge. As a function of the chemical potential, $\sigma^{\mathrm{LPL}}_{zxy}$ can change its sign, indicating that the direction of the induced magnetization can be switched by adjusting the chemical potential. Let us estimate the magnitude of the induced magnetization in this model. We assume a typical relaxation time for metals, $\tau\approx10^{-14}\,\mathrm{s}$, a lattice constant $a=2\times10^{-10}\,\mathrm{m}$, a photon energy $\hbar\omega=0.1\,\mathrm{eV}$, and an electric-field amplitude of $10^7\,\mathrm{V/m}$. With these parameters, the orders of magnitudes of the magnetizations induced by CPL and LPL are estimated to be $10^{-14}$ and $10^{-13}\,\mathrm{A}$, respectively. These values are of the same order of magnitude as, or even larger than, those predicted for other theoretically proposed mechanisms, such as the inverse Faraday effect in metals~\cite{Hertel_2006_IFE} and disordered metals~\cite{Taguchi_Tatara_2011_IFE}, and the orbital inverse Faraday and inverse Cotton--Mouton effects in Hall fluids~\cite{cardosoOrbitalInverseFaraday2025}. Our results, Eqs.~\eqref{eq:sigma_LPL} and \eqref{eq:sigma_CPL}, can be distinguished from these proposed mechanisms by their distinct dependence on the relaxation time $\tau$ and the frequency $\omega$. The inverse Faraday effects discussed in Refs.~\cite{Hertel_2006_IFE,cardosoOrbitalInverseFaraday2025} scale as $\tau^0\omega^{-3}$ in the high frequency regime, whereas the inverse Faraday effect discussed in Ref.~\cite{Taguchi_Tatara_2011_IFE} scales as $\tau^2\omega$. These $\tau$- and $\omega$-dependencies differ from those obtained here and therefore provide a clear means to distinguish mechanisms of different physical origin.

\section{Conclusion}
\label{sec:concl}

In this work, we have developed a general theory of light-induced magnetization in electronic systems originating from quantum-geometric effects. By considering spatially nonuniform electric fields within a semiclassical framework based on the Boltzmann transport equation, we established a unified formalism in which the magnetization arises as a second-order response to the electric field.

We identified two distinct quantum-geometric contributions to the magnetization: the quantum metric quadrupole and weighted quantum metric. While the former has been previously discussed in the context of transport currents under nonuniform electric field, the latter originates from perturbations to the Bloch states and does not appear in conventional transport currents. Our results demonstrate that both contributions are essential for a complete description of light-induced magnetization and that they naturally give rise to responses under both LPL and CPL. These effects correspond to the inverse Cotton--Mouton and the inverse Faraday effects, respectively.

We further analyzed symmetry constraints in two-dimensional systems and clarified the conditions under which responses to LPL and CPL are allowed. Using both a continuum massive Dirac model with anisotropic quadratic corrections and a tight-binding model on the hexagonal lattice, we explicitly demonstrated the emergence of nonvanishing magnetization responses when either mirror or rotational symmetries are broken. In particular, the tight-binding model of anisotropic gapped graphene allowed us to estimate the magnitude of the effect and showed that the induced magnetization is experimentally observable, although it may not constitute the dominant contribution to the overall inverse Faraday and inverse Cotton--Mouton effects.

Our work establishes optically generated magnetization as a direct and experimentally accessible manifestation of quantum geometry in solids. Beyond providing a unified description of magneto-optical effects, the present framework opens new avenues for probing quantum-geometric quantities through nonlinear optical measurements.

\section*{Acknowledgements}
We thank Teruaki Suyama, So Tanaka, and Ryutaro Tomomatsu for useful discussions and comments. HY is supported by Japan Society for the Promotion of Science (JSPS) KAKENHI Grant Number JP24KJ1109 and by MEXT Initiative to Establish Next-generation Novel Integrated Circuits Centers (X-NICS) Grant Number JPJ011438. TY is supported by Japan Society for the Promotion of Science (JSPS) KAKENHI Grant Numbers JP30578216.

\section*{Data availability}
The data that support the findings of this paper are available upon reasonable request.

\begin{widetext}
\appendix

\section{Equation of motion}
\label{app:EOM}
Here we calculate each term in the equation of motion Eqs.~\eqref{eq:EOM} up to the order of $E\partial E$.

First, as $\Omega'^{qr}$ contains $\partial E$ and $E\partial E$, $\dot{q}_a$ reads
\be
    \dot{q}_a&=\frac{Q}{\hbar}E_a(\bm{r},t)+\Omega'^{rq}_{ab}\dot{q}_b\nonumber\\
    &=\frac{Q}{\hbar}E_a(\bm{r},t)-\dot{q}_b\Omega'^{qr}_{ba}\nonumber\\
    &=\frac{Q}{\hbar}E_a(\bm{r},t)-\left(\frac{Q}{\hbar}E_b(\bm{r},t)-\dot{q}_c\Omega'^{qr}_{cb}\right)\Omega'^{qr}_{ba}\nonumber\\
    &=\frac{Q}{\hbar}E_a(\bm{r},t)-\frac{Q}{\hbar}E_b(\bm{r},t)\Omega'^{qr}_{ba}+\mathcal{O}\left(E^2\right)\nonumber\\
    &\approx \frac{Q}{\hbar}E_a+\frac{1}{2}\left(\frac{Q}{\hbar}\right)^2G_{bj}E_b\partial_{r_a}E_j.\label{eq:dotq_EdE}
\ee
We next calculate $\dot{r}_a$. It contains perturbed energy $\varepsilon_n'$ and the Berry curvature in the momentum space $\Omega'^{qq}_{ab}$.

For $\varepsilon_n'$, the procedure of the calculation is the same as in Ref.~\cite{GaoXiao_2019_NonreciprocalDirectional}. We write $\varepsilon'_n=\varepsilon_n+\delta\varepsilon$ and calculate $\delta\varepsilon$, which contains terms up to the order of $E\partial E$. From the perturbed state Eq.~\eqref{eq:PerturbedState}
\be
    \left\lvert u_n'(\bm{r},\bm{q})\right\rangle &=|u_n\rangle+\sum_{m\neq n}\frac{QE_a(\bm{r},t)(A_{mn})_a}{\varepsilon_m-\varepsilon_n}|u_m\rangle,\nonumber
\ee
we construct a wave-packet as
\be
    \left\lvert W_n\right\rangle &\coloneqq \int\left[\rmd\bm{q}\right]C_n(\bm{q})e^{i\bm{q}\cdot\bm{r}}\left\lvert u'_n\right\rangle.
\ee
Here, the expansion coefficient $C_n$ is localized around the center of mass momentum of the wave-packet $\bm{q}_c$ and approximated as $\abs{C_n}^2=(2\pi)^d\delta(\bm{q}-\bm{q}_c)$. The center of mass of the wave-packet in real space is defined as
\be
    \bm{r}_c&\coloneqq \left\langle W_n\right\rvert\hat{\bm{r}}\left\lvert W_n\right\rangle.
\ee
Around this point, we expand the scalar potential as
\be
    \phi(\bm{r})&=\phi(\bm{r}_c)-Q\bm{E}\cdot(\bm{r}-\bm{r}_c)-\frac{Q}{2}\partial_{r_i}E_j\left(r_i-r_{c,i}\right)\left(r_j-r_{c,j}\right)+\cdots.
\ee
Then, the correction to the energy is given by
\be
    \delta\varepsilon &= \left\langle W_n\right\rvert\left(-Q\bm{E}\cdot(\hat{\bm{r}}-\bm{r}_c)-\frac{Q}{2}\partial_{r_i}E_j\left(\hat{r}_i-r_{c,i}\right)\left(\hat{r}_j-r_{c,j}\right)\right)\left\lvert W_n\right\rangle\nonumber\\
    &=-\frac{Q}{2}\partial_{r_i}E_j\left\langle W_n\right\rvert\left(\hat{r}_i-r_{c,i}\right)\left(\hat{r}_j-r_{c,j}\right)\left\lvert W_n\right\rangle\nonumber\\
    &=-\frac{Q}{2}\partial_{r_i}E_j\left(\left\langle W_n\right\rvert\hat{r}_i\hat{r}_j\left\lvert W_n\right\rangle-r_{c,i}r_{c,j}\right).
\ee
We first calculate the center of position up to the order of $E$.
\be
    r_{c,i}&=\left\langle W_n\right\rvert\hat{r}_i\left\lvert W_n\right\rangle\nonumber\\
    &=\int \rmd\bm{r}\left\langle W_n\middle|\bm{r}\right\rangle\left\langle\bm{r}\right\rvert\hat{r}_i\left\lvert W_n\right\rangle\nonumber\\
    &=\int \rmd\bm{r}\int\left[\rmd\bm{q}\right]\int\left[\rmd\bm{q}'\right] C^*_n(\bm{q}')e^{-i\bm{q}'\cdot\bm{r}}\left\langle u'_n(\bm{q}')\middle|\bm{r}\right\rangle r_iC_n(\bm{q})e^{i\bm{q}\cdot\bm{r}}\left\langle\bm{r}\middle| u'_n(\bm{q})\right\rangle\nonumber\\
    &=\int \rmd\bm{r}\int\left[\rmd\bm{q}\right]\int\left[\rmd\bm{q}'\right] C^*_n(\bm{q}')C_n(\bm{q})e^{-i\bm{q}'\cdot\bm{r}}\left\langle u'_n(\bm{q}')\middle|\bm{r}\right\rangle \left(-i\frac{\partial}{\partial q_i}e^{i\bm{q}\cdot\bm{r}}\right)\left\langle\bm{r}\middle| u'_n(\bm{q})\right\rangle\nonumber\\
    &=i\int \rmd\bm{r}\int\left[\rmd\bm{q}\right]\int\left[\rmd\bm{q}'\right] C^*_n(\bm{q}')e^{-i(\bm{q}'-\bm{q})\cdot\bm{r}}\left\langle u'_n(\bm{q}')\middle|\bm{r}\right\rangle \frac{\partial}{\partial q_i}\left(C_n(\bm{q})\left\langle\bm{r}\middle| u'_n(\bm{q})\right\rangle\right)\nonumber\\
    &=i\int \rmd\bm{r}\int\left[\rmd\bm{q}\right]\int\left[\rmd\bm{q}'\right] C^*_n(\bm{q}')e^{-i(\bm{q}'-\bm{q})\cdot\bm{r}}\left\{u'^{*}_{n,q'}(\bm{r})u'_{n,q}(\bm{r})\frac{\partial C_n(\bm{q})}{\partial q_i}+C_n(\bm{q})u'^{*}_{n,q'}(\bm{r})\frac{\partial u'_{n,q}(\bm{r})}{\partial q_i}\right\}\label{eq:rc_middle}
\ee
We have the orthonormality of Bloch states
\be
    \left\langle \psi_m(\bm{q}')\middle| \psi_n(\bm{q})\right\rangle&=\int\rmd\bm{r}\left\langle \psi_m(\bm{q}')\middle|\bm{r}\right\rangle \left\langle\bm{r}\middle|\psi_n(\bm{q})\right\rangle\nonumber\\
    &=\int\rmd\bm{r}e^{-i(\bm{q'}-\bm{q})\cdot\bm{r}}u^*_{m,q'}(\bm{r})u_{n,q}(\bm{r})\nonumber\\
    &=\sum_{\bm{R}}\int_{\mathrm{UC}}\rmd\bm{r}e^{-i(\bm{q'}-\bm{q})\cdot(\bm{r}+\bm{R})}u^*_{m,q'}(\bm{r})u_{n,q}(\bm{r})\qquad (\because u_{n,q}(\bm{r}+\bm{R})=u_{n,q}(\bm{r}))\nonumber\\
    &=(2\pi)^d\delta(\bm{q}'-\bm{q})\int\rmd \bm{r}u^*_{m,q}(\bm{r})u_{n,q}(\bm{r})\nonumber\\
    &=(2\pi)^d\delta(\bm{q}'-\bm{q})\delta_{nm},
\ee
where we used the periodicity of the periodic part of the Bloch state to reduce the integration over the whole space to the unit cell and the sum over all unit cells. By this relation, the first term in Eq.~\eqref{eq:rc_middle} is approximated to be products of unperturbed states in the approximation up to $E$. For the second term, we can ignore the position dependence of $E(\bm{r})$ in the perturbed state and thus, $u'_{n,q}(\bm{r})$ is also periodic in this approximation. Then, by a similar transformation as above,
\be
    r_{c,i}&=i\int \rmd\bm{r}\int\left[\rmd\bm{q}\right]\int\left[\rmd\bm{q}'\right] C^*_n(\bm{q}')e^{-i(\bm{q}'-\bm{q})\cdot\bm{r}}\left\{u'^{*}_{n,q'}(\bm{r})u'_{n,q}(\bm{r})\frac{\partial C_n(\bm{q})}{\partial q_i}+C_n(\bm{q})u'^{*}_{n,q'}(\bm{r})\frac{\partial u'_{n,q}(\bm{r})}{\partial q_i}\right\}\nonumber\\
    &=i\int\left[\rmd\bm{q}\right]\int\left[\rmd\bm{q}'\right]\sum_{\bm{R}}e^{-i(\bm{q}'-\bm{q})\cdot\bm{R}}\int_{\mathrm{UC}} \rmd\bm{r} C^*_n(\bm{q}')e^{-i(\bm{q}'-\bm{q})\cdot\bm{r}}\left\{u^{*}_{n,q'}(\bm{r})u_{n,q}(\bm{r})\frac{\partial C_n(\bm{q})}{\partial q_i}+C_n(\bm{q})u'^{*}_{n,q'}(\bm{r})\frac{\partial u'_{n,q}(\bm{r})}{\partial q_i}\right\}\nonumber\\
    &\hspace{15cm}+\mathcal{O}(E^2,\partial E)\nonumber\\
    &\approx i\int\left[\rmd\bm{q}\right]\int_{\mathrm{UC}} \rmd\bm{r} C^*_n(\bm{q})\left\{u^{*}_{n,q}(\bm{r})u_{n,q}(\bm{r})\frac{\partial C_n(\bm{q})}{\partial q_i}+C_n(\bm{q})u'^{*}_{n,q}(\bm{r})\frac{\partial u'_{n,q}(\bm{r})}{\partial q_i}\right\}.\nonumber
\ee
When we write $C_n(\bm{q})=\abs{C_n}e^{-i\gamma(\bm{q})}$, we obtain
\be
    r_{c,i}&=\int\left[\rmd\bm{q}\right]\abs{C_n}^2\left(\partial_{q_i}\gamma+\left\langle u'_n\middle|i\partial_{q_i}u'_n\right\rangle\right)\nonumber\\
    &=\left(\partial_{q_i}\gamma+\left\langle u'_n\middle|i\partial_{q_i}u'_n\right\rangle\right)\eval_{\bm{q}=\bm{q}_c}.
\ee
For the dipole moment term, a similar calculation can be performed as
\be
    \left\langle W_n\right\rvert\hat{r}_i\hat{r}_j\left\lvert W_n\right\rangle&=\int \rmd\bm{r}\left\langle W_n\middle|\bm{r}\right\rangle\left\langle\bm{r}\right\rvert\hat{r}_i\hat{r}_j\left\lvert W_n\right\rangle\nonumber\\
    &=\int \rmd\bm{r}\int\left[\rmd\bm{q}\right]\int\left[\rmd\bm{q}'\right] C^*_n(\bm{q}')C_n(\bm{q})\left(i\frac{\partial}{\partial q'_i}e^{-i\bm{q}'\cdot\bm{r}}\right)\left\langle u'_n(\bm{q}')\middle|\bm{r}\right\rangle \left(-i\frac{\partial}{\partial q_j}e^{i\bm{q}\cdot\bm{r}}\right)\left\langle\bm{r}\middle| u'_n(\bm{q})\right\rangle\nonumber\\
    &=\int \rmd\bm{r}\int\left[\rmd\bm{q}\right]\int\left[\rmd\bm{q}'\right]e^{-i(\bm{q}'-\bm{q})\cdot\bm{r}}\left(i\frac{\partial}{\partial q'_i}\left(C^*_n(\bm{q}')u'^*_{n,q'}(\bm{r})\right)\right) \left(-i\frac{\partial}{\partial q_j}\left(C_n(\bm{q})u'_{n,q}(\bm{r})\right)\right)\nonumber\\
    &\approx\int_{\mathrm{UC}}\rmd\bm{r}\int\left[\rmd\bm{q}\right]\left(i\frac{\partial}{\partial q_i}\left(C^*_n(\bm{q})u'^*_{n,q}(\bm{r})\right)\right) \left(-i\frac{\partial}{\partial q_j}\left(C_n(\bm{q})u'_{n,q}(\bm{r})\right)\right)\nonumber\\
    &=\int_{\mathrm{UC}}\rmd\bm{r}\int\left[\rmd\bm{q}\right] \abs{C_n}^2\left(\frac{\partial u'^*_{n,q}}{\partial q_i}\frac{\partial u'_{n,q}}{\partial q_j}+i u'^*_{n,q}\partial_{q_i}\gamma\frac{\partial u'_{n,q}}{\partial q_j}-iu'_{n,q}\partial_{q_j}\gamma\frac{\partial u'^*_{n,q}}{\partial q_i}+\partial_{q_i}\gamma \partial_{q_j}\gamma u'^*_{n,q}u'_{n,q}\right)\nonumber\\
    &\approx \left(\left\langle \partial_{q_i} u'_n\middle|\partial_{q_j}u'_n\right\rangle+i\partial_{q_i}\gamma\left\langle u'_n\middle|i\partial_{q_j}u'_n\right\rangle-\partial_{q_j}\gamma\left\langle i\partial_{q_i}u'_n\middle|u'_n\right\rangle+\partial_{q_i}\gamma \partial_{q_j}\gamma\right)\eval_{\bm{q}=\bm{q}_c}.
\ee
Therefore, we obtain
\be
    \left\langle W_n\right\rvert\hat{r}_i\hat{r}_j\left\lvert W_n\right\rangle-r_{c,i}r_{c,j}&=\left\langle \partial_{q_i} u'_n\middle|\partial_{q_j}u'_n\right\rangle+i\partial_{q_i}\gamma\left\langle u'_n\middle|i\partial_{q_j}u'_n\right\rangle-\partial_{q_j}\gamma\left\langle i\partial_{q_i}u'_n\middle|u'_n\right\rangle+\partial_{q_i}\gamma \partial_{q_j}\gamma\nonumber\\
    &\qquad -\left(\partial_{q_i}\gamma+\left\langle u'_n\middle|i\partial_{q_i}u'_n\right\rangle\right)\left(\partial_{q_j}\gamma+\left\langle u'_n\middle|i\partial_{q_j}u'_n\right\rangle\right)\nonumber\\
    &=\left\langle \partial_{q_i} u'_n\middle|\partial_{q_j}u'_n\right\rangle-\left\langle \partial_{q_i} u'_n\middle|u'_n\right\rangle\left\langle u'_n\middle|\partial_{q_j}u'_n\right\rangle\nonumber\\
    &=\partial_i\left(\langle u_n|+\sum_{m\neq n}\frac{QE_a(\bm{r},t)(A_{nm})_a}{\varepsilon_m-\varepsilon_n}\langle u_m|\right)\partial_j\left(|u_n\rangle+\sum_{m\neq n}\frac{QE_a(\bm{r},t)(A_{mn})_a}{\varepsilon_m-\varepsilon_n}|u_m\rangle\right)\nonumber\\
    &\quad -\partial_i\left(\langle u_n|+\sum_{m\neq n}\frac{QE_a(\bm{r},t)(A_{nm})_a}{\varepsilon_m-\varepsilon_n}\langle u_m|\right)\left(|u_n\rangle+\sum_{m\neq n}\frac{QE_a(\bm{r},t)(A_{mn})_a}{\varepsilon_m-\varepsilon_n}|u_m\rangle\right)\nonumber\\
    &\qquad\times    \left(\langle u_n|+\sum_{m\neq n}\frac{QE_a(\bm{r},t)(A_{nm})_a}{\varepsilon_m-\varepsilon_n}\langle u_m|\right)\partial_j\left(|u_n\rangle+\sum_{m\neq n}\frac{QE_a(\bm{r},t)(A_{mn})_a}{\varepsilon_m-\varepsilon_n}|u_m\rangle\right)\nonumber\\
    &=\left\langle \partial_i u_n\middle|\partial_j u_n\right\rangle-\left\langle \partial_i u_n\middle|u_n\right\rangle\left\langle u_n\middle|\partial_j u_n\right\rangle\nonumber\\
    &\qquad +QE_a\sum_{m\neq n}\left[\partial_i\left(\frac{(A_{nm})_a}{\varepsilon_m-\varepsilon_n}\langle u_m|\right)|\partial_j u_n\rangle+\langle \partial_i u_n|\partial_j\left(\frac{(A_{mn})_a}{\varepsilon_m-\varepsilon_n}|u_m\rangle\right)\right]\nonumber\\
    &\qquad -QE_a\sum_{m\neq n}\left[\partial_i\left(\frac{(A_{nm})_a}{\varepsilon_m-\varepsilon_n}\langle u_m|\right)|u_n\rangle\langle u_n|\partial_j u_n\rangle+\langle \partial_i u_n|\left(\frac{(A_{mn})_a}{\varepsilon_m-\varepsilon_n}|u_m\rangle\right)\left\langle u_n\middle|\partial_j u_n\right\rangle\right.\nonumber\\
    &\hspace{2cm} \left.+\langle \partial_i u_n|u_n\rangle \left(\frac{(A_{nm})_a}{\varepsilon_m-\varepsilon_n}\langle u_m|\right) \left\lvert\partial_j u_n\right\rangle+\langle \partial_i u_n|u_n\rangle\langle u_n|\partial_j \left(\frac{(A_{mn})_a}{\varepsilon_m-\varepsilon_n}|u_m\rangle\right)\right]\nonumber\\
    &=g_{ij}+QE_a\sum_{\substack{m\neq n\\l\neq n}}\left[\partial_i\left(\frac{(A_{nm})_a}{\varepsilon_m-\varepsilon_n}\langle u_m|\right)|u_l\rangle\langle u_l|\partial_j u_n\rangle+\langle \partial_i u_n|u_l\rangle\langle u_l|\partial_j \left(\frac{(A_{mn})_a}{\varepsilon_m-\varepsilon_n}|u_m\rangle\right)\right]\nonumber\\
    &\qquad -QE_a\sum_{m\neq n}\left[\frac{(A_{mn})_a}{\varepsilon_m-\varepsilon_n}(A_{nm})_i(A_{nn})_j+\frac{(A_{nm})_a}{\varepsilon_m-\varepsilon_n}(A_{nn})_i(A_{mn})_j\right]\nonumber\\
    &=g_{ij}-QE_a\sum_{m\neq n}\frac{(A_{mn})_a}{\varepsilon_m-\varepsilon_n}\left[\langle\partial_i \partial_j u_n|u_m\rangle+\langle \partial_i u_n|\partial_j u_m\rangle+\sum_{l\neq n}\langle\partial_i u_n|u_l\rangle\langle \partial_j u_l| u_m\rangle\right]\nonumber\\
    &\qquad -QE_a\sum_{m\neq n}\frac{(A_{nm})_a}{\varepsilon_m-\varepsilon_n}\left[\langle u_m|\partial_i \partial_j u_n\rangle + \langle \partial_i u_m|\partial_j u_n\rangle+\sum_{l\neq n}\langle u_m| \partial_i u_l\rangle \langle u_l|\partial_j u_n\rangle\right]\nonumber\\
    &\qquad -QE_a\sum_{m\neq n}\left[\frac{(A_{mn})_a}{\varepsilon_m-\varepsilon_n}(A_{nm})_i(A_{nn})_j+\frac{(A_{nm})_a}{\varepsilon_m-\varepsilon_n}(A_{nn})_i(A_{mn})_j\right]\nonumber\\
    &=g_{ij}-QE_a\sum_{m\neq n}\frac{(A_{mn})_a}{\varepsilon_m-\varepsilon_n}\left[\langle\partial_i \partial_j u_n|u_m\rangle+(A_{nn})_i(A_{nm})_j\right]\nonumber\\
    &\qquad -QE_a\sum_{m\neq n}\frac{(A_{nm})_a}{\varepsilon_m-\varepsilon_n}\left[\langle u_m|\partial_i \partial_j u_n\rangle + (A_{mn})_i(A_{nn})_j\right]\nonumber\\
    &\qquad -QE_a\sum_{m\neq n}\left[\frac{(A_{mn})_a}{\varepsilon_m-\varepsilon_n}(A_{nm})_i(A_{nn})_j+\frac{(A_{nm})_a}{\varepsilon_m-\varepsilon_n}(A_{nn})_i(A_{mn})_j\right]\nonumber\\
    &=g_{ij}-2QE_a\sum_{m\neq n}\operatorname{Re}\left[\frac{(A_{nm})_a}{\varepsilon_m-\varepsilon_n}\left\{\langle u_m|\partial_i \partial_j u_n\rangle + (A_{nn})_i(A_{mn})_j + (A_{mn})_i(A_{nn})_j\right\}\right].
\ee
This quantity is invariant under a gauge transformation $|u_n\rangle\to e^{i\phi_n(\bm{q})}|u_n\rangle$. Then, the energy correction up to the order of $E\partial E$ is given as
\be
    \delta\varepsilon &= -\frac{Q}{2}\partial_{r_i}E_j\left(g_{ij}-2QE_a\sum_{m\neq n}\operatorname{Re}\left[\frac{(A_{nm})_a}{\varepsilon_m-\varepsilon_n}\left\{\langle u_m|\partial_i \partial_j u_n\rangle + (A_{nn})_i(A_{mn})_j + (A_{mn})_i(A_{nn})_j\right\}\right]\right)
\ee

From Eqs.~\eqref{eq:def_Omegaqq} and \eqref{eq:PerturbedState}, the Berry curvature is
\be
    \Omega'^{qq}_{ab}&=\Omega^{qq}_{ab}-2QE_i\operatorname{Im}\left[\sum_{m\neq n}\frac{\partial}{\partial q_a}\left(\frac{\left(A_{nm}\right)_i}{\varepsilon_m-\varepsilon_n}\left\langle u_m\right\rvert \right)\left\lvert\frac{\partial u_n}{\partial q_b}\right\rangle +\left\langle\frac{\partial u_n}{\partial q_b}\right\rvert\frac{\partial}{\partial q_b}\left(\frac{\left(A_{mn}\right)_i}{\varepsilon_m-\varepsilon_n}\left\lvert u_m\right\rangle\right)\right]+\mathcal{O}(E^2).
\ee
Then, using these results, $\dot{r}_a$ up to the order of $E\partial E$ is given by
\be
    \dot{r}_a&=\frac{1}{\hbar}\frac{\partial\varepsilon'_n}{\partial q_a}-\Omega'^{qq}_{ab}\dot{q}_b-\Omega'^{qr}_{ab}\dot{r}_b\nonumber\\
    &\approx \frac{1}{\hbar}\frac{\partial\varepsilon_n}{\partial q_a}-\frac{Q}{2\hbar}\frac{\partial g_{ij}}{\partial q_a}\partial_{r_i}E_j+\frac{Q^2}{\hbar}\frac{\partial}{\partial q_a}\sum_{m\neq n}\operatorname{Re}\left[\frac{(A_{nm})_b}{\varepsilon_m-\varepsilon_n}\left\{\langle u_m|\partial_i \partial_j u_n\rangle + (A_{nn})_i(A_{mn})_j + (A_{mn})_i(A_{nn})_j\right\}\right]E_b\partial_{r_i}E_j\nonumber\\
    &\qquad -\left(\Omega^{qq}_{ab}-2QE_i\operatorname{Im}\left[\sum_{m\neq n}\frac{\partial}{\partial q_a}\left(\frac{\left(A_{nm}\right)_i}{\varepsilon_m-\varepsilon_n}\left\langle u_m\right\rvert \right)\left\lvert\frac{\partial u_n}{\partial q_b}\right\rangle +\left\langle\frac{\partial u_n}{\partial q_b}\right\rvert\frac{\partial}{\partial q_b}\left(\frac{\left(A_{mn}\right)_i}{\varepsilon_m-\varepsilon_n}\left\lvert u_m\right\rangle\right)\right]\right)\nonumber\\
    &\hspace{10cm}\times\left(\frac{Q}{\hbar}E_a+\frac{1}{2}\left(\frac{Q}{\hbar}\right)^2G_{bj}E_b\partial_{r_a}E_j\right)\nonumber\\
    &\qquad -\left(-\frac{Q}{2\hbar}G_{aj}\partial_{r_b}E_j-2Q^2\Im\left[\sum_{m\neq n}\partial_{q_a}\left(\frac{\left(A_{nm}\right)_i}{\varepsilon_m-\varepsilon_n}\right)\frac{\left(A_{mn}\right)_j}{\varepsilon_m-\varepsilon_n}+i\sum_{\substack{m\neq n\\l\neq n}}\frac{\left(A_{nm}\right)_i\left(A_{ml}\right)_a\left(A_{ln}\right)_j}{(\varepsilon_m-\varepsilon_n)(\varepsilon_l-\varepsilon_n)}\right]E_i\partial_{r_b}E_j\right)\dot{r}_b\nonumber\\
    &\approx \frac{1}{\hbar}\frac{\varepsilon_n}{\partial q_a}-\frac{Q}{\hbar}\Omega^{qq}_{ab}E_b+\left(-\frac{Q}{2\hbar}\frac{\partial g_{ij}}{\partial q_a}\partial_{r_i}E_j+\frac{Q}{2\hbar^2}G_{aj}\frac{\partial \varepsilon_b}{\partial q_b}\partial_{r_b}E_j\right)\nonumber\\
    &\qquad +\left(-\frac{Q^2}{2\hbar^2}G_{aj}\Omega^{qq}_{bc}E_c\partial_{r_b}E_j-\frac{Q^2}{2\hbar^2}G_{ij}\Omega^{qq}_{ab}E_i\partial_{r_b}E_j\right.\nonumber\\
    &\quad\left.+\frac{2Q^2}{\hbar}\frac{\partial \varepsilon_n}{\partial q_b}\Im\left[\sum_{m\neq n}\partial_{q_a}\left(\frac{\left(A_{nm}\right)_i}{\varepsilon_m-\varepsilon_n}\right)\frac{\left(A_{mn}\right)_j}{\varepsilon_m-\varepsilon_n}+i\sum_{\substack{m\neq n\\l\neq n}}\frac{\left(A_{nm}\right)_i\left(A_{ml}\right)_a\left(A_{ln}\right)_j}{(\varepsilon_m-\varepsilon_n)(\varepsilon_l-\varepsilon_n)}\right]E_i\partial_{r_b}E_j\right.\nonumber\\
    &\qquad \left.+\frac{Q^2}{\hbar}\frac{\partial}{\partial q_a}\sum_{m\neq n}\operatorname{Re}\left[\frac{(A_{nm})_b}{\varepsilon_m-\varepsilon_n}\left\{\langle u_m|\partial_i \partial_j u_n\rangle + (A_{nn})_i(A_{mn})_j + (A_{mn})_i(A_{nn})_j\right\}\right]E_b\partial_{r_i}E_j\right).
\ee

\section{Distribution function}
\label{app:distribution function}
We write the distribution function by a Fourier series up to the terms containing $E\partial E$.

We write the distribution function $f(\bm{q},t)$ using a complex representation as
\be
    f(\bm{q},t)&=\Re\left[\sum_{n=0}^{\infty}\tilde{f}_n(\bm{q})e^{in\omega t}\right]\qquad \left(\tilde{f}_n\in\mathbb{C}\right).\label{eq:f_expansion}
\ee
The Boltzmann equation~\eqref{eq:Boltzmann} gives
\be
    f(\bm{q},t)&=f_0(\bm{q})-\left(\tau \dot{q}_a\frac{\partial}{\partial q_a}+\tau\frac{\partial}{\partial t}\right)f(\bm{q},t),
\ee
where $f_0(\bm{q})$ is an equilibrium distribution function, $E_a=\Re\left[\tilde{E}_ae^{i\omega t}\right]$ is the electric field, and $\tau$ is the relaxation time. By substituting Eq.~\eqref{eq:f_expansion}, we obtain
\be
    \sum_{n=0}^{\infty}\tilde{f}_n(\bm{q})e^{in\omega t}+\mathrm{c.c.}&=\left[f_0-\tau \dot{q}_a\sum_{n=0}^{\infty}\frac{\partial \tilde{f}_n}{\partial q_a}e^{in\omega t}-\tau\sum_{n=0}^{\infty}in\omega \tilde{f}_ne^{in\omega t}\right]+\mathrm{c.c.}.
\ee
We want to obtain $\tilde{f}_n$ neglecting terms with $E^2$ and higher orders of $E$. We employ a result Eq.~\eqref{eq:dotq_EdE} for $\dot{q}_a$:
\be
    \dot{q}_a&\approx\frac{Q}{2\hbar}\left(\tilde{E}_ae^{i\omega t}+\tilde{E}_a^*e^{-i\omega t}\right)E_a+\frac{1}{2}\left(\frac{Q}{2\hbar}\right)^2G_{bj}\left(\tilde{E}_be^{i\omega t}+\tilde{E}_b^*e^{-i\omega t}\right)\partial_{r_a}\left(\tilde{E}_je^{i\omega t}+\tilde{E}_j^*e^{-i\omega t}\right).\nonumber
\ee
Then, we obtain
\be
    \sum_{n=0}^{\infty}\tilde{f}_n(\bm{q})e^{in\omega t}+\mathrm{c.c.}&\approx\left[f_0-\tau \left(\frac{Q}{2\hbar}\left(\tilde{E}_ae^{i\omega t}+\tilde{E}_a^*e^{-i\omega t}\right)E_a+\frac{1}{2}\left(\frac{Q}{2\hbar}\right)^2G_{bj}\left(\tilde{E}_be^{i\omega t}+\tilde{E}_b^*e^{-i\omega t}\right)\partial_{r_a}\left(\tilde{E}_je^{i\omega t}+\tilde{E}_j^*e^{-i\omega t}\right)\right)\right.\nonumber\\
    &\qquad\left.\times\sum_{n=0}^{\infty}\frac{\partial \tilde{f}_n}{\partial q_a}e^{in\omega t}-\tau\sum_{n=0}^{\infty}in\omega \tilde{f}_ne^{in\omega t}\right]+\mathrm{c.c.}\nonumber\\
    &=\left[f_0-\tau\frac{Q}{2\hbar}\left(\tilde{E}_a\sum_{n=0}^{\infty}\frac{\partial \tilde{f}_n}{\partial q_a}e^{i(n+1)\omega t}+\tilde{E}_a^*\sum_{n=0}^{\infty}\frac{\partial \tilde{f}_n}{\partial q_a}e^{i(n-1)\omega t}\right)\right.\nonumber\\
    &\qquad \left.+\tau\frac{1}{2}\left(\frac{Q}{2\hbar}\right)^2G_{bj}\left(\tilde{E}_b\partial_a\tilde{E}_j^*+\tilde{E}_b\partial_a\tilde{E}_je^{2i\omega t}+\mathrm{c.c.}\right)\sum_{n=0}^{\infty}\frac{\partial \tilde{f}_n}{\partial q_a}e^{in\omega t}-\tau\sum_{n=0}^{\infty}in\omega \tilde{f}_ne^{in\omega t}\right]+\mathrm{c.c.}\nonumber\\
    &=\left[f_0+\sum_{n=0}^{\infty}\left\{\frac{\tau}{2}\left(\frac{Q}{2\hbar}\right)^2G_{bj}\left(\tilde{E}_b\partial_a\tilde{E}_j^*+\mathrm{c.c.}\right)\frac{\partial \tilde{f}_n}{\partial q_a}-in\tau\omega\tilde{f}_n\right\}e^{in\omega t}\right.\nonumber\\
    &\qquad-\frac{\tau Q}{2\hbar}\left(\tilde{E}_a\sum_{n=1}^{\infty}\frac{\partial \tilde{f}_{n-1}}{\partial q_a}e^{in\omega t}+\tilde{E}_a^*\sum_{n=-1}^{\infty}\frac{\partial \tilde{f}_{n+1}}{\partial q_a}e^{in\omega t}\right)\nonumber\\
    &\qquad\left.+\frac{\tau}{2}\left(\frac{Q}{2\hbar}\right)^2G_{bj}\left(\tilde{E}_b\partial\tilde{E}_j\sum_{n=2}^{\infty}\frac{\partial \tilde{f}_{n-2}}{\partial q_a}e^{in\omega t}+\tilde{E}_b^*\partial\tilde{E}_j^*\sum_{n=-2}^{\infty}\frac{\partial \tilde{f}_{n+2}}{\partial q_a}e^{in\omega t}\right)\right]+\mathrm{c.c.}
\ee
By equating terms at each order of $e^{i\omega t}$, we obtain
\be
    \tilde{f}_0&=f_0+\frac{\tau}{2}\left(\frac{\tau Q}{2\hbar}\right)^2G_{bj}\tilde{E}_b\partial_a\tilde{E}_j^*\frac{\partial \left(\tilde{f}_0+\tilde{f}_0^*\right)}{\partial q_a}-\frac{\tau Q}{2\hbar}\tilde{E}_a^*\frac{\partial \tilde{f}_{1}^*}{\partial q_a}++\frac{\tau}{2}\left(\frac{\tau Q}{2\hbar}\right)^2G_{bj}\tilde{E}_b^*\partial_a\tilde{E}_j^*\frac{\partial \tilde{f}_2}{\partial q_a},\\
    \tilde{f}_1&=-\frac{\tau Q}{2\hbar}\tilde{E}_a\frac{\partial \left(\tilde{f}_0+\tilde{f}_0^*\right)}{\partial q_a}+\frac{\tau}{2}\left(\frac{\tau Q}{2\hbar}\right)^2G_{bj}\left[\tilde{E}_b\partial_a\tilde{E}_j^*\frac{\partial \left(\tilde{f}_1+\tilde{f}_1^*\right)}{\partial q_a}+\tilde{E}_b\partial_a\tilde{E}_j\frac{\partial \tilde{f}_1^*}{\partial q_a}\right]-i\tau\omega\tilde{f}_1\nonumber\\
    &\hspace{5cm}-\frac{\tau Q}{2\hbar}\tilde{E}_a^*\frac{\partial \tilde{f}_2}{\partial q_a}+\frac{\tau}{2}\left(\frac{\tau Q}{2\hbar}\right)^2G_{bj}\tilde{E}_b^*\partial_a\tilde{E}_j^*\frac{\partial \tilde{f}_3}{\partial q_a},\\
    \tilde{f}_2&=\frac{\tau}{2}\left(\frac{\tau Q}{2\hbar}\right)^2G_{bj}\tilde{E}_b\partial_a\tilde{E}_j\frac{\partial \left(\tilde{f}_0+\tilde{f}_0^*\right)}{\partial q_a}-\frac{\tau Q}{2\hbar}\tilde{E}_a\frac{\partial \tilde{f}_1}{\partial q_a}+\frac{\tau}{2}\left(\frac{\tau Q}{2\hbar}\right)^2G_{bj}\tilde{E}_b\partial_a\tilde{E}_j^*\frac{\partial \left(\tilde{f}_2+\tilde{f}_2^*\right)}{\partial q_a}-2i\tau\omega\tilde{f}_2\nonumber\\
    &\hspace{5cm}-\frac{\tau Q}{2\hbar}\tilde{E}_a^*\frac{\partial \tilde{f}_3}{\partial q_a}+\frac{\tau}{2}\left(\frac{\tau Q}{2\hbar}\right)^2G_{bj}\tilde{E}_b^*\partial_a\tilde{E}_j^*\frac{\partial \tilde{f}_4}{\partial q_a},\\
    \tilde{f}_n&=\frac{\tau}{2}\left(\frac{\tau Q}{2\hbar}\right)^2G_{bj}\tilde{E}_b\partial_a\tilde{E}_j^*\frac{\partial\left(\tilde{f}_n+\tilde{f}_n^*\right)}{\partial q_a}-in\tau\omega\tilde{f}_n\nonumber\\
    &\quad -\frac{\tau Q}{2\hbar}\left(\tilde{E}_a\frac{\partial \tilde{f}_{n-1}}{\partial q_a}+\tilde{E}_a^*\frac{\partial \tilde{f}_{n+1}}{\partial q_a}\right)+\frac{\tau}{2}\left(\frac{\tau Q}{2\hbar}\right)^2G_{bj}\left(\tilde{E}_b\partial_a\tilde{E}_j\frac{\partial\tilde{f}_{n-2}}{\partial q_a}+\tilde{E}_b^*\partial_a\tilde{E}_j^*\frac{\partial\tilde{f}_{n+2}}{\partial q_a}\right)\qquad (n\geq 3).
\ee
By using these relationships, we can obtain all $\tilde{f}_n$. As $\tilde{f}_n$ contains at least $\tilde{E}^n$, up to second-order of the electric field, we obtain 
\be
    \tilde{f}_0&\approx f_0+\tau\left(\frac{\tau Q}{2\hbar}\right)^2G_{bj}\tilde{E}_b\partial_a\tilde{E}_j^*\frac{\partial \tilde{f}_0}{\partial q_a},\\
    \tilde{f}_1&\approx -\frac{\tau Q}{\hbar}\tilde{E}_a\frac{\partial \tilde{f}_0}{\partial q_a}-i\tau\omega\tilde{f}_1,\\
    \tilde{f}_2&\approx\tau\left(\frac{\tau Q}{2\hbar}\right)^2G_{bj}\tilde{E}_b\partial_a\tilde{E}_j\frac{\partial \tilde{f}_0}{\partial q_a}-2i\tau\omega\tilde{f}_2,\\
    \tilde{f}_n&\approx 0\qquad (n\geq 3).
\ee
Therefore, we obtain distribution function including the effect of nonuniform electric field up to $E\partial E$ as Eqs.~\eqref{eq:f0}-\eqref{eq:fn}.

\section{Polarization of light}
\label{app:polarization}

To make this paper self-contained, here we briefly summarize the relationship between polarization of light and second-order responses of its electric field.

Without loss of generality, we consider an electromagnetic wave propagating in the $z$-direction with frequency $\omega$. The, the electric field can be written as
\be
    \bm{E}&=\operatorname{Re}\left[\tilde{\bm{E}}e^{i(kz-\omega t)}\right],
\ee
where a complex vector
\be
    \tilde{\bm{E}}&=\frac{E_0}{\sqrt{a^2+b^2}}\left(\begin{matrix}
        a\\be^{i\theta}\\0
    \end{matrix}\right)e^{i\phi}
\ee
determines the polarization of the light. Here, $E_0,a,b,\theta,\phi$ are all real values. When $\theta=n\pi\ (n\in\mathbb{Z})$, $x$ and $y$ components oscillate without any phase difference and represent LPL. On the other hand, for $\theta\neq n\pi$, the light is elliptically polarized. When $\theta=(2n+1)\pi/2$, polarization is perfectly circular.

Let us consider the following products of two electric fields:
\be
    \tilde{E}_x\tilde{E}_x^*&=\frac{a^2}{a^2+b^2}E_0^2,\nonumber\\
    \tilde{E}_y\tilde{E}_y^*&=\frac{b^2}{a^2+b^2}E_0^2,\nonumber\\
    \tilde{E}_x\tilde{E}_y^*+\tilde{E}_y\tilde{E}_x^*&=\frac{ab}{a^2+b^2}E_0^2\cdot2\cos\theta,\nonumber\\
    \tilde{E}_x\tilde{E}_y^*-\tilde{E}_y\tilde{E}_x^*&=\frac{ab}{a^2+b^2}E_0^2\cdot\left(-2i\right)\sin\theta.\nonumber
\ee
Therefore, for LPL (i.e. $\theta=n\pi$), $\operatorname{Im}\left[\tilde{E}_x\tilde{E}_y^*\right]=0$ and for CPL (i.e. $\theta=(2n+1)\pi/2$), $\operatorname{Re}\left[\tilde{E}_x\tilde{E}_y^*\right]=0$.

Assume that a physical quantity $\bm{X}$ is expressed in complex representation as $\bm{X}=\operatorname{Re}\left[\tilde{\bm{X}}\right]$ and it is a second order of the complex electric field,
\be
    \tilde{X}_a&=\alpha_{abc}\tilde{E}_b\tilde{E}_c^*.
\ee
Then, the physical quantity can be calculated by taking the real part of both sides as
\be
    X_a&=\operatorname{Re}\left[\alpha_{abc}\tilde{E}_b\tilde{E}_c^*\right]\nonumber\\
    &=\operatorname{Re}\left[\alpha_{abc}\right]\operatorname{Re}\left[\tilde{E}_b\tilde{E}_c^*\right]-\operatorname{Im}\left[\alpha_{abc}\right]\operatorname{Im}\left[\tilde{E}_b\tilde{E}_c^*\right].
\ee
When we consider LPL, the real part of the coefficient $\alpha_{abc}$ contribute to the final response. Furthermore, since $\operatorname{Re}\left[\tilde{E}_b\tilde{E}_c^*\right]$ is symmetric with respect to the exchange of indices $b$ and $c$, only the symmetric part of $\alpha_{abc}$ contribute to the response to LPL. It is now natural to define
\be
    \alpha^{\mathrm{LPL}}_{abc}&\coloneqq \operatorname{Re}\left[\frac{\alpha_{abc}+\alpha_{acb}}{2}\right]=\alpha^{\mathrm{LPL}}_{acb}.
\ee
Similarly, for CPL, only the imaginary and antisymmetric part of the coefficient contribute to the physical phenomena. Thus, we define
\be
    \alpha^{\mathrm{CPL}}_{abc}&\coloneqq -\operatorname{Im}\left[\frac{\alpha_{abc}-\alpha_{acb}}{2}\right]=-\alpha^{\mathrm{CPL}}_{acb}.
\ee
Using these quantities, we can decompose $X$ to two parts, representing responses to LPL and CPL as
\be
    X_a&=\frac{\operatorname{Re}\left[\alpha_{abc}\right]\operatorname{Re}\left[\tilde{E}_b\tilde{E}_c^*\right]+\operatorname{Re}\left[\alpha_{acb}\right]\operatorname{Re}\left[\tilde{E}_c\tilde{E}_b^*\right]}{2}-\frac{\operatorname{Im}\left[\alpha_{abc}\right]\operatorname{Im}\left[\tilde{E}_b\tilde{E}_c^*\right]+\operatorname{Im}\left[\alpha_{acb}\right]\operatorname{Im}\left[\tilde{E}_c\tilde{E}_b^*\right]}{2}\nonumber\\
    &=\alpha^{\mathrm{LPL}}_{abc}\operatorname{Re}\left[\tilde{E}_b\tilde{E}_c^*\right]+\alpha^{\mathrm{CPL}}_{abc}\operatorname{Im}\left[\tilde{E}_b\tilde{E}_c^*\right].
\ee

\section{Restrictions by crystalline symmetries}
\label{app:sym}
We discuss the restrictions imposed by the mirror symmetry and rotation symmetry of the system on the magnetization induced by second-order of electric field. In our formulation, magnetization can be induced only in the $z$-direction and thus, we focus on the response
\be
    \tilde{M}^0_z&=\sigma_{zab}\tilde{E}_a\tilde{E}^*_b\qquad(a,b\in\left\{x,y\right\}).\label{eq:symmetry_base}
\ee

\subsection{Mirror symmetry}
First of all, symmetry with respect to the mirror plane perpendicular to the $z$-axis does not impose any restrictions as this mirror operations keeps the form of Eq.~\eqref{eq:symmetry_base}. Next, we consider a mirror operation $m_{\perp}$ whose mirror plane is perpendicular to the system. Suppose that the angle between this mirror plane and the $x$-axis is $\phi$. Then, the electric field is transformed as
\be
    \tilde{\bm{E}}=\left(\begin{matrix}
        \tilde{E}_x\\\tilde{E}_y
    \end{matrix}\right)\to\left(\begin{matrix}
        \cos2\phi&\sin2\phi\\\sin2\phi&-\cos2\phi
    \end{matrix}\right)\left(\begin{matrix}
        \tilde{E}_x\\\tilde{E}_y
    \end{matrix}\right).
\ee
On the other hand, $m_{\perp}$ changes the sign of $\tilde{M}^0_z$. Therefore, this operation transforms Eq.~\eqref{eq:symmetry_base} as
\be
    \tilde{M}^0_z&=\tilde{\bm{E}}^T\sigma_z\tilde{\bm{E}}^*\nonumber\\
    \to -\tilde{M}^0_z&=\tilde{\bm{E}}^T\left(\begin{matrix}
        \cos2\phi&\sin2\phi\\\sin2\phi&-\cos2\phi
    \end{matrix}\right)^T\sigma_z\left(\begin{matrix}
        \cos2\phi&\sin2\phi\\\sin2\phi&-\cos2\phi
    \end{matrix}\right)\tilde{\bm{E}}^*.
\ee
Then, the invariance of the system under this symmetry imposes the condition for $\sigma_z:=\left(\begin{matrix}
    \sigma_{zxx}&\sigma_{zxy}\\
    \sigma_{zyx}&\sigma_{zyy}
\end{matrix}\right)$ as
\be
    \sigma_z&=-\left(\begin{matrix}
        \cos2\phi&\sin2\phi\\\sin2\phi&-\cos2\phi
    \end{matrix}\right)\sigma_z\left(\begin{matrix}
        \cos2\phi&\sin2\phi\\\sin2\phi&-\cos2\phi
    \end{matrix}\right)\nonumber\\
    &=-\left(
        \begin{array}{cc}
        \sigma_{zxx}C^2+\sigma_{zyy}S^2+(\sigma_{zxy}+\sigma_{zyx})SC&\left(\sigma_{zxx}-\sigma_{zyy}\right)SC-\sigma_{zxy}C^2+\sigma_{zyx}S^2\\
        \left(\sigma_{zxx}-\sigma_{zyy}\right)SC+\sigma_{zxy}S^2-\sigma_{zyx}C^2&\sigma_{zxx}S^2+\sigma_{zyy}C^2-(\sigma_{zxy}+\sigma_{zyx})SC\\
        \end{array}
        \right)\nonumber\\
    \Leftrightarrow&\left(
        \begin{array}{cc}
        \left(\sigma_{zxx}+\sigma_{zyy}\right)+\left(\sigma_{zxx}-\sigma_{zyy}\right)C^2+(\sigma_{zxy}+\sigma_{zyx})SC&\left(\sigma_{zxx}-\sigma_{zyy}\right)SC+\left(\sigma_{zxy}+\sigma_{zyx}\right)S^2\\
        \left(\sigma_{zxx}-\sigma_{zyy}\right)SC+\left(\sigma_{zxy}+\sigma_{zyx}\right)S^2&\left(\sigma_{zxx}+\sigma_{zyy}\right)-\left(\sigma_{zxx}-\sigma_{zyy}\right)C^2-(\sigma_{zxy}+\sigma_{zyx})SC
        \end{array}
        \right)=0,
\ee
where we used shorthand notations $C:=\cos2\phi$ and $S:=\sin2\phi$. This indicates that
\be
    \sin2\phi\neq0,\cos2\phi\neq0&\Rightarrow\sigma_{zxx}=-\sigma_{zyy},\sigma_{zyy}=\frac{\sigma_{zxy}+\sigma_{zyx}}{2}\tan2\phi,\\
    \sin2\phi\neq0,\cos2\phi=0&\Rightarrow\sigma_{zxx}=-\sigma_{zyy},\sigma_{zxy}=-\sigma_{zyx},\\
    \sin2\phi=0,\cos2\phi\neq0&\Rightarrow\sigma_{zxx}=\sigma_{zyy}=0.
\ee
Therefore, if the system has $m_{100}$ or $m_{010}$ symmetry, $\sigma_{zaa}=0\ (a=x,y)$. If the mirror symmetry is $m_{110}$ or $m_{1\bar{1}0}$, then $\sigma^{\mathrm{LPL}}_{zxy}=0$ and $\sigma_{zxx}=-\sigma_{zyy}$. Otherwise, the existence of the mirror symmetry requires $\sigma_{zxx}=-\sigma_{zyy}$ and relationship among remaining three components but does not force them to be zero.

\subsection{Rotation symmetry}
Next we consider the effect of rotation symmetry around an axis perpendicular to the system. Rotation of an angle $\theta$ around the $z$-axis changes the electric field as
\be
    \tilde{\bm{E}}=\left(\begin{matrix}
        \tilde{E}_x\\\tilde{E}_y
    \end{matrix}\right)\to\left(\begin{matrix}
        \cos\theta&-\sin\theta\\\sin\theta&\cos\theta
    \end{matrix}\right)\left(\begin{matrix}
        \tilde{E}_x\\\tilde{E}_y
    \end{matrix}\right).
\ee
In contrast to the mirror operation with respect to the plane perpendicular to the system, rotation operation around the $z$-axis keeps $\tilde{M}^0_z$ unchanged. Thus, the invariance under this rotation operation requires
\be
    \sigma_z&=\left(\begin{matrix}
        \cos\theta&-\sin\theta\\\sin\theta&\cos\theta
    \end{matrix}\right)^T\sigma_z\left(\begin{matrix}
        \cos\theta&-\sin\theta\\\sin\theta&\cos\theta
    \end{matrix}\right)\nonumber\\
    \Leftrightarrow&\left(\begin{matrix}
        \left(\sigma_{zxx}-\sigma_{zyy}\right)S^2+\left(\sigma_{zxy}+\sigma_{zyx}\right)SC&\left(\sigma_{zxx}-\sigma_{zyy}\right)SC+\left(\sigma_{zxy}+\sigma_{zyx}\right)S^2\\
        \left(\sigma_{zxx}-\sigma_{zyy}\right)SC+\left(\sigma_{zxy}+\sigma_{zyx}\right)S^2&-\left(\sigma_{zxx}-\sigma_{zyy}\right)S^2+\left(\sigma_{zxy}+\sigma_{zyx}\right)SC
    \end{matrix}\right)=0,
\ee
where $S:=\sin\theta$ and $C:=\cos\theta$. This indicates that
\be
    \sin\theta\neq0,\cos\theta\neq0&\Rightarrow \sigma_{zxx}=\sigma_{zyy},\sigma_{zxy}=-\sigma_{zyx},\\
    \sin\theta\neq0,\cos\theta=0&\Rightarrow \sigma_{zxx}=\sigma_{zyy},\sigma_{zxy}=-\sigma_{zyx},\\
    \sin\theta=0,\cos\theta\neq0&\Rightarrow \mathrm{Always\ satisfied.}
\ee
Therefore, $C_{2z}$-symmetry of the system does not constrain the response. Other rotation symmetries force $\sigma^{\mathrm{LPL}}_{zxy}=0$ and $\sigma_{zxx}=\sigma_{zyy}$.

\subsection{Rotation and mirror symmetry}
Combining two results discussed above, in the presence of both a mirror symmetry with respect to a plane perpendicular to the system $m_{\perp}$ and a rotation symmetry $C_{nz}\ (n\geq3)$, response to LPL is forbidden. On the other hand, these crystalline symmetries do not pose any conditions for the response to CPL.

\section{Quantum metric quadrupole}
\label{app:QMQ}
Here we present expressions of quantum metric quadrupole and other terms in Eqs.~\eqref{eq:sigma_LPL} and \eqref{eq:sigma_CPL} for general two-level systems. Then, we apply them to our model system.

In general, Hamiltonian of a two-level system can be written as
\be
    H(\bm{q})&=h_0(\bm{q})\sigma_0+\bm{h}(\bm{q})\cdot\boldsymbol{\sigma},\label{eq:app_H_twoband}
\ee
where $\sigma_0$ is an identity matrix and $\boldsymbol{\sigma}$ is a vector of Pauli matrices. Then, for the lower band, quantum metric can be given as
\be
    g_{ab}&=\operatorname{Re}\left[\left(A_{01}\right)_a\left(A_{10}\right)_b\right]\nonumber\\
    &=\frac{1}{4h^2}\left(\partial_a h_i\partial_b h_i-\partial_a h\partial_b h\right),
\ee
where $h=\abs{\bm{h}}$, $\partial_a=\partial/\partial q_a,\, (a=x,y,z)$, and the sum over repeated indices is taken. The quantum metric dipole density and quantum metric quadrupole density are
\be
    \partial_cg_{ab}&=-\frac{\partial_c h}{2h^3}(\partial_{a}h_i\partial_{b}h_i-\partial_{a}h\partial_{b}h)+\frac{1}{4h^2}\partial_c(\partial_{a}h_i\partial_{b}h_i-\partial_{a}h\partial_{b}h),\\
    \partial_{d}\partial_{c}g_{ab}&=\frac{3\partial_{c}h\partial_{d}h}{2h^4}\qty(\partial_{a}h_i\partial_{b}h_i-\partial_{a}h\partial_{b}h)-\frac{\partial_{d}\partial_{c}h}{2h^3}\qty(\partial_{a}h_i\partial_{b}h_i-\partial_{a}h\partial_{b}h)\nonumber\\
    &\quad -\frac{\partial_{c}h}{2h^3}\partial_{d}\qty(\partial_{a}h_i\partial_{b}h_i-\partial_{a}h\partial_{b}h)-\frac{\partial_{d}h}{2h^3}\partial_{c}\qty(\partial_{a}h_i\partial_{b}h_i-\partial_{a}h\partial_{b}h)+\frac{1}{4h^2}\partial_{c}\partial_{d}\qty(\partial_{a}h_i\partial_{b}h_i-\partial_{a}h\partial_{b}h).
\ee
For the calculation of $\sigma^{\mathrm{LPL}}$ and $\sigma^{\mathrm{CPL}}$, we need an anti-symmetric part of the quantum metric quadrupole;
\be
    \varepsilon_{\mu ca}\partial_{d}\partial_{c}g_{ab}&=\varepsilon_{\mu ca}\left[\frac{3\partial_{c}h\partial_{d}h}{2h^4}\partial_{a}h_i\partial_{b}h_i-\frac{\partial_{d}\partial_{c}h}{2h^3}\partial_{a}h_i\partial_{b}h_i-\frac{\partial_{c}h}{2h^3}\partial_{d}\left(\partial_{a}h_i\partial_{b}h_i\right)\right.\nonumber\\
    &\hspace{5cm}\left.-\frac{\partial_{d}h}{2h^3}\qty(\partial_{a}h_i\partial_{c}\partial_{b}h_i-\partial_{a}h\partial_{c}\partial_{b}h)+\frac{1}{4h^2}\partial_d\qty(\partial_{a}h_i\partial_{c}\partial_{b}h_i-\partial_{a}h\partial_{c}\partial_{b}h)\right]\label{eq:app_quadrupole}.
\ee
The symmetric part of Eq.~\eqref{eq:app_quadrupole} with respect to the exchange of indices $d\leftrightarrow b$ contributes to $\sigma^{\mathrm{LPL}}$ and the anti-symmetric part to $\sigma^{\mathrm{CPL}}$.

For the $\sigma^{\mathrm{CPL}}$, we also need a term containing the weighted quantum metric. In a two-band system described by Eq.~\eqref{eq:app_H_twoband}, we have $G_{ab}=2\hbar g_{ab}/h$. Thus,
\be
    \partial_d\left(G_{ab}v_c\right)&=2\left(\partial_dg_{ab}\frac{\partial_c\left(h_0-h\right)}{h}-g_{ab}\frac{\partial_dh\partial_c\left(h_0-h\right)}{h^2}+g_{ab}\frac{\partial_d\partial_c\left(h_0-h\right)}{h}\right).
\ee

\section{Anisotropic massive Dirac model}
\label{app:anisotropic_massive_dirac}

Let us assume a Hamiltonian of the form
\be
    H(\bm{q})&=\alpha_xq_x\sigma_x+\alpha_yq_y\sigma_y+M\sigma_z,\label{eq:Anisotropic_Dirac_Hamiltonian}
\ee
which is an effective model of an anisotropic, massive, and two-dimensional Dirac cone. For this system, the quantum metric, metric dipole density, and metric quadrupole density are
\be
    g_{ab}&=\frac{\alpha_a^2}{4h^2}\left(\delta_{ab}-\alpha_b^2\frac{q_aq_b}{h^2}\right),\\
    \partial_cg_{ab}&=-\frac{1}{4h^2}\left(\alpha_a^2\alpha_b^2\delta_{bc}q_a+\alpha_a^2\alpha_b^2\delta_{ca}q_b+\alpha_b^2\alpha_c^2\delta_{ab}q_c+\alpha_a^2\alpha_c^2\delta_{ab}q_c\right)+\alpha_a^2\alpha_b^2\alpha_c^2\frac{q_aq_bq_c}{h^6},\\
    \partial_d\partial_cg_{ab}&=-\frac{1}{4h^2}\left(\alpha_a^2\alpha_b^2\delta_{bc}\delta_{ad}+\alpha_a^2\alpha_b^2\delta_{ca}\delta_{bd}+\alpha_b^2\alpha_c^2\delta_{ab}\delta_{cd}+\alpha_a^2\alpha_c^2\delta_{ab}\delta_{cd}\right)\nonumber\\
    &\quad +\frac{\alpha_d^2}{h^6}\left[\alpha_a^2\alpha_b^2\left(\delta_{bc}q_aq_d+\delta_{ac}q_bq_d+\delta_{cd}q_aq_b\right)+\alpha_b^2\alpha_c^2\left(\delta_{ab}q_cq_d+\delta_{ad}q_bq_c\right)+\alpha_a^2\alpha_c^2\left(\delta_{ab}q_cq_d+\delta_{bd}q_aq_c\right)\right]\nonumber\\
    &\qquad -6\alpha_a^2\alpha_b^2\alpha_c^2\alpha_d^2\frac{q_aq_bq_cq_d}{h^8},
\ee
where $h=\sqrt{(\alpha_xq_x)^2+(\alpha_yq_y)^2+M^2}$. Similarly,
\be
    G_{ab}v_c&=-\frac{\alpha_a\alpha_c}{2h^4}q_c\left(\delta_{ab}-\alpha_b^2\frac{q_aq_b}{h^2}\right),\\
    \partial_d\left(G_{ab}v_c\right)&=-\frac{\alpha_a^2\alpha_c^2}{2h^4}\delta_{ab}\delta_{cd}+\frac{\alpha_a^2\alpha_c^2}{2h^6}\left(4\alpha_d^2\delta_{ab}q_cq_d+\alpha_b^2\left(\delta_{ad}q_bq_c+\delta_{bd}q_aq_c+\delta_{cd}q_aq_b\right)\right)-3\alpha_a^2\alpha_b^2\alpha_c^2\alpha_d^2\frac{q_aq_bq_cq_d}{h^8}.
\ee
Using these quantities, we obtain
\be
    \sigma^{\mathrm{LPL}}_{zxx}&\propto\int\left[\rmd\bm{q}\right]f_0\left[-\frac{10 \alpha_x^4 \alpha_y^2 q_x q_y}{\left(\alpha_x^2 q_x^2+\alpha_y^2 q_y^2+M^2\right)^3}\right],\\
    \sigma^{\mathrm{LPL}}_{zxy}&\propto\int\left[\rmd\bm{q}\right]f_0\left[\frac{5 \alpha_x^2 \alpha_y^2}{\left(\alpha_x^2 q_x^2+\alpha_y^2 q_y^2+M^2\right)^3}\left(\alpha_x^2q_x^2-\alpha_y^2q_y^2\right)\right],\\
    \sigma^{\mathrm{LPL}}_{zyy}&\propto\int\left[\rmd\bm{q}\right]f_0\left[\frac{10 \alpha_x^2 \alpha_y^4 q_x q_y}{\left(\alpha_x^2 q_x^2+\alpha_y^2 q_y^2+M^2\right)^3}\right],\\
    \sigma^{\mathrm{CPL}}_{zxy}&\propto\int\left[\rmd\bm{q}\right]f_0\left[\frac{5 \alpha_x^2 \alpha_y^2 \left(\alpha_x^2 q_x^2+\alpha_y^2 q_y^2-M^2\right)}{2 \left(\alpha_x^2 q_x^2+\alpha_y^2 q_y^2+M^2\right)^3}\right].
\ee
By introducing new variables $u:=\alpha_xq_x$ and $v:=\alpha_yq_y$, the integrands of $\sigma^{\mathrm{LPL}}_{zxx}$ and $\sigma^{\mathrm{LPL}}_{zyy}$ are odd functions of $u$ and $v$ and thus, these integrations give zeros. Similarly, the integrand of $\sigma^{\mathrm{LPL}}_{zxy}$ is odd under the exchange of $u$ and $v$, indicating $\sigma^{\mathrm{LPL}}_{zxy}=0$. Therefore, $\sigma^{\mathrm{LPL}}_{zab}=0$ for the simplest massive Dirac model.

\end{widetext}

\bibliography{IFE.bib}

\end{document}